\documentclass[%
 reprint,
 superscriptaddress,
 amsmath,amssymb,
 aps,
 showkeys,
]{revtex4-2}

\usepackage{graphicx}% Include figure files
\usepackage{dcolumn}% Align table columns on decimal point
\usepackage{bm}% bold math
\usepackage{multirow}
\usepackage{booktabs}
\usepackage{float}
\usepackage{placeins}
\usepackage{xcolor} %from Sandra - delete again if you prefer
\usepackage[colorlinks=true,linkcolor=blue,citecolor=blue,urlcolor=blue]{hyperref}% add hypertext capabilities
\usepackage{miller}

\begin{document}

\preprint{APS/123-QED}

% Main title of the paper
\title{On the temperature and chemical dependency of prismatic stacking faults in C14 Laves phases} 

\author{Zhuocheng Xie}
\email[]{xie@imm.rwth-aachen.de}
\affiliation{Institute of Physical Metallurgy and Materials Physics, RWTH Aachen University, 52056 Aachen, Germany}

\author{Dimitri Chauraud}
\affiliation{Max-Planck-Institut für Eisenforschung GmbH, Max-Planck-Str. 1, 40237 Düsseldorf, Germany}
     
\author{Erik Bitzek}
\affiliation{Max-Planck-Institut für Eisenforschung GmbH, Max-Planck-Str. 1, 40237 Düsseldorf, Germany}

\author{Sandra Korte-Kerzel}
\affiliation{Institute of Physical Metallurgy and Materials Physics, RWTH Aachen University, 52056 Aachen, Germany}

\author{Julien Guénolé}
\email[]{julien.guenole@univ-lorraine.fr}
\affiliation{Université de Lorraine, CNRS, Arts et Métiers ParisTech, LEM3, 57070 Metz, France}
\affiliation{Labex Damas, Université de Lorraine, 57070 Metz, France}

\begin{abstract}
Activation of non-basal slip is essential in improving the deformability of hexagonal crystals. However, the mechanism of non-basal slip remains largely unknown, especially for complex intermetallics such as Laves phases. In this work, the prismatic slip systems of C14 Laves crystals and possible metastable states along the slip paths are assessed using atomistic simulations. 
Multiple prismatic stacking fault states with the same lattice pattern but different site occupancies and chemical distributions are identified at different inter-atomic layers, temperatures and chemical compositions. The formation of energetically favorable prismatic stacking faults involves short-range diffusion which implies the thermally activated nature of prismatic slip. 
The outcomes of this work advance the understanding of temperature and chemical-dependent non-basal slip in Laves phases and can be extended to other topologically close-packed phases.

\end{abstract}

\maketitle

Laves phases are intermetallics that are present in various alloys and exert a significant influence on their mechanical characteristics due to their high strength and superior creep resistance in comparison to the matrix phases \cite{sinha1972topologically,paufler2011early,stein2021laves}. Despite their excellent mechanical performance at elevated temperatures, Laves phase alloys are commonly deemed unsuitable for structural applications because of their well-known brittleness at room temperature \cite{stein2021laves,livingston1992laves,pollock2010weight}.
The crystal structure of Laves phases with an ideal chemical composition AB\textsubscript{2} consists of a triple-layer and a kagomé layer along the basal ($\{ 1 1 1 \}$) planes. Synchro-shear \cite{kronberg1957plastic,chisholm2005dislocations}, as the most energetically favorable mechanism of plasticity on basal or $\{ 1 1 1 \}$ planes in Laves phases, is widely studied, particularly using atomic-scale modeling \cite{vedmedenko2008first,guenole2019basal,xie2022unveiling,xie2023thermally}. However, the mechanisms underlying plastic deformation on non-basal (non-$\{ 1 1 1 \}$) planes in Laves phases are not well understood so far.

In the early 1970s, Paufler et al. \cite{paufler1970mobility,kubsch1974mobility} reported the activation of grown-in prismatic dislocations in the prototype C14 MgZn\textsubscript{2} under uniaxial compression at temperatures around 400\textdegree C.
Recently, Zehnder et al. \cite{zehnder2019plastic} investigated the plastic deformation of single crystal C14 CaMg\textsubscript{2} at room temperature by nanoindentation and microcompression. The prismatic slip was confirmed by slip trace analysis and exhibits the lowest critical resolved shear stress (CRSS) among basal and non-basal slip systems. The authors \cite{freund2021plastic} further investigated the activation of these slip systems of the CaMg\textsubscript{2} phase in the temperature range from 50 to 250\textdegree C, where prismatic slip shows a decreasing CRSS with temperature.
Zhang et al. \cite{zhang2020shuffle} characterized two prismatic stacking fault (SF) structures with the same crystal lattice but different site occupancies in a C14 Nb-based Laves phase using atomic-resolution scanning transmission electron microscopy (STEM) after deformation at different temperature regimes.
The formation of high and room-temperature prismatic SFs was interpreted based on the morphology of prismatic planes and topologically close-packed rules. The high-temperature SF formation was attributed to the diffusion-assisted shuffle mechanism, while the low-temperature SF formation was associated with the glide mechanism.
In addition, the prismatic SFs were not flat and exhibited numerous steps, which was attributed to dislocation motion through frequent switching between different prismatic slip planes.

So far, the deformation mechanisms of prismatic plasticity in C14 Laves phases, particularly the processes that lead to the formation of prismatic SFs through partial dislocations, are still not well understood. In this study, we investigated the temperature and chemical-dependent prismatic slip in C14 CaMg\textsubscript{2} and MgZn\textsubscript{2} Laves phases by performing atomistic simulations using the semi-empirical potentials \cite{kim2015modified,brommer2009vibrational}. The minimum energy paths (MEPs) of slip events on different prismatic interlayers and corresponding prismatic SF states were identified using the nudged elastic band (NEB) method \cite{henkelman2000climbing,henkelman2000improved}. The effects of temperature and chemical composition on SF structures were studied using molecular dynamics (MD) and hybrid MD/Monte Carlo (MC) simulations, respectively. The FIRE \cite{bitzek2006structural,guenole2020assessment} and Quickmin \cite{sheppard2008optimization} algorithms were used for relaxation in molecular statics (MS) simulations and NEB calculations, respectively. For more details on the simulation setup and methods, see Section I in the Supplementary Material.

\begin{figure*}[htbp!]
\centering
\includegraphics[width=\textwidth]{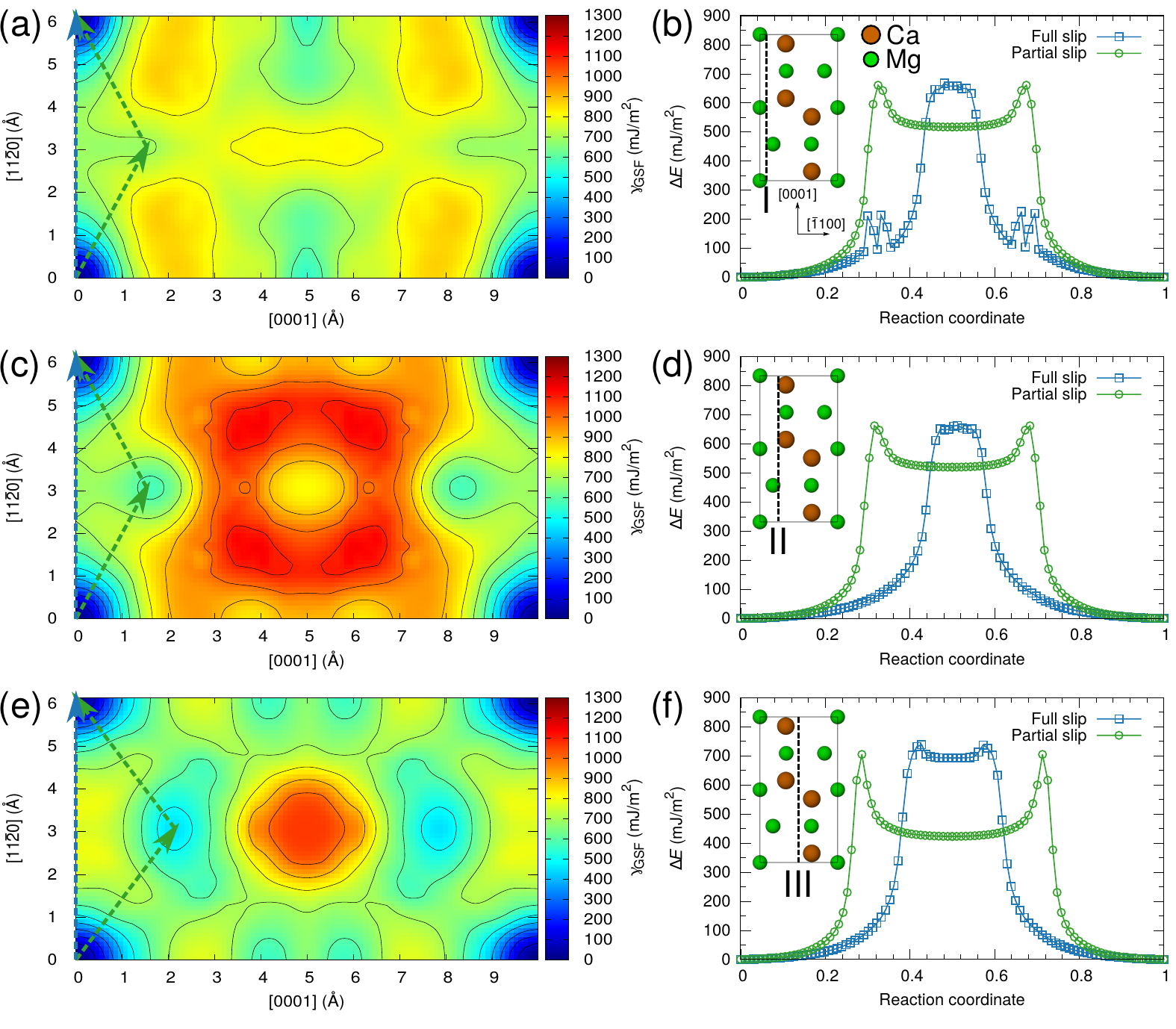}
\caption{Prismatic slip systems of C14 CaMg\textsubscript{2}. Generalized stacking fault energy ($\gamma$) surfaces of interlayer (a) I, (c) II and (e) III. Minimum energy paths computed for full ($[1 1 \bar{2} 0]$ or $\langle a \rangle$) and partial ($\langle a+c \rangle$) slip on interlayer (b) I, (d) II and (f) III using NEB. The paths of full and partial slip are marked in blue and green dashed arrows on the $\gamma$ surfaces, respectively. The C14 unit cell and prismatic slip planes (marked in black dashed lines) are shown in the insets. Ca (large) and Mg (small) atoms are colored brown and green, respectively.}
\label{fig1}
\end{figure*}

The prismatic slip systems were systematically assessed by calculating the generalized stacking fault energy ($\gamma$) surfaces of three possible interlayers on prismatic $\{\bar{1} 1 0 0\}$ plane in C14 CaMg\textsubscript{2} (see Figure \ref{fig1}(a,c,e)) and MgZn\textsubscript{2} (see Figure \ref{figS1}(a,c,e)). Meta-stable states close to the $[1 1 \bar{2} 0]$ slip path were identified as the prismatic SF states in all interlayers. After full relaxation, the SF of interlayer III (SF III) shows the lowest stacking fault energy (SFE: 422 mJ/m\textsuperscript{2} in C14 CaMg\textsubscript{2} and 265 mJ/m\textsuperscript{2} in C14 MgZn\textsubscript{2}), presumably is more stable than the SF states of interlayer I and II (SF I and SF II). The energy barriers of full and partial slip paths were calculated using the NEB method. Although the interplanar distance of interlayer III is almost two times larger than interlayer I and II, the energy barriers of slip events on the three interlayers are at similar levels. For the full sip, the activation energies on interlayer III are 11 \% and 3 \% higher than on interlayer I and II in C14 CaMg\textsubscript{2} and for MgZn\textsubscript{2}, respectively. In C14 CaMg\textsubscript{2}, the activation energy of partial slip on interlayer III is 7 \% higher than interlayer I and II. For the partial slip in MgZn\textsubscript{2}, the activation energy on interlayer III becomes 12 \% lower than interlayer I and II counterparts. The energy barrier for partial slip within the same interlayer is slightly lower compared to that of full slip.

\begin{table*}[!htbp]
\centering
\caption[]{\label{tab.1}Summary of interplaner distances of prismatic interlayers, activation energies of full ($\Delta E$\textsubscript{full}) and partial slip ($\Delta E$\textsubscript{partial}) on prismatic planes and prismatic stacking fault energies (SFE) of C14 CaMg\textsubscript{2} and MgZn\textsubscript{2} and SFE of optimized prismatic stacking fault structures.}
\centering
\scriptsize
\begin{tabular}{p{0.2\textwidth}p{0.05\textwidth}p{0.05\textwidth}p{0.05\textwidth}p{0.05\textwidth}p{0.05\textwidth}p{0.05\textwidth}}
\hline\hline
\addlinespace[0.1cm]
\multicolumn{1}{l}{} & \multicolumn{3}{c}{CaMg\textsubscript{2}} & \multicolumn{3}{c}{MgZn\textsubscript{2}}\\
\cmidrule(lr){2-4}
\cmidrule(lr){5-7}
Interlayer & I & II & III & I & II & III \\
\addlinespace[0.1cm]
\hline
\addlinespace[0.1cm]
Interplanar distance (\AA) & 0.902 & 0.871 & 1.712 & 0.760 & 0.731 & 1.433 \\
$\Delta E$\textsubscript{full} (mJ/m\textsuperscript{2}) & 667 & 664 & 738 & 495 & 495 & 512 \\
$\Delta E$\textsubscript{partial} (mJ/m\textsuperscript{2}) & 661 & 662 & 705 & 476 & 476 & 420 \\
SFE (mJ/m\textsuperscript{2}) & 516 & 516 & 422 & 460 & 460 & 265 \\
Optimized SFE (mJ/m\textsuperscript{2}) & 374 & 374 & 326 & 191 & 191 & 189 \\
\addlinespace[0.1cm]
\hline\hline
\end{tabular}
\end{table*}

The identification of prismatic SF states is in agreement with the experimental observation in C14 Nb-base Laves phase after low and high-temperature deformation by Zhang et al. \cite{zhang2020shuffle}. Abundant extended prismatic SFs bounded by partial dislocations with compact cores and Burgers vectors of $\frac{3}{16}[0 0 0 1 ]$ + $\frac{1}{6}[1 1 \bar{2} 0 ]$ were observed in atomic-resolution STEM \cite{zhang2020shuffle}. Our study determined the Burgers vectors of prismatic partial slip close to $\frac{3}{16}[0 0 0 1]$ + $\frac{1}{6}[1 1 \bar{2} 0]$, which correlates well with the above-mentioned experimental results. Furthermore, Zhang et al. \cite{zhang2020shuffle} reported that the prismatic SF is not flat and exhibits numerous steps, which they attributed to the movement of dislocations by frequently switching on different prismatic slip planes. This can be explained by nearly equal energy barriers of prismatic slip on interlayer I, II, and III, see Table \ref{tab.1}. Further research is needed to gain a more comprehensive understanding of the out-of-plane mechanisms of prismatic dislocations switching between various interlayers and the associated activation energies. 

\begin{figure*}[htbp!]
\centering
\includegraphics[width=0.75\textwidth]{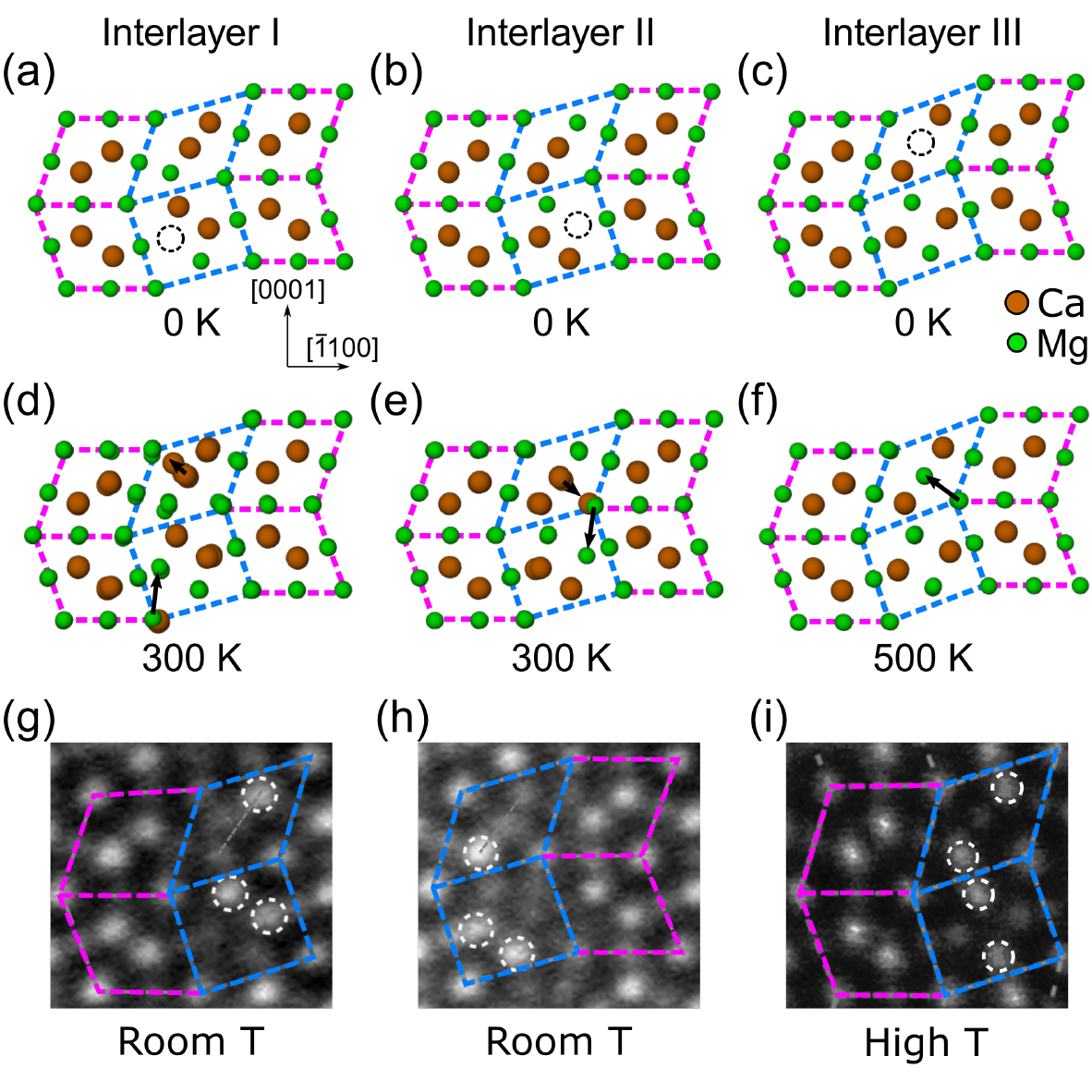}
\caption{Prismatic stacking faults in C14 CaMg\textsubscript{2}. Stacking faults of interlayer (a) I, (b) II and (c) III after energy minimization (force tolerance: $10^{-8}$ eV/\AA). Stacking fault configurations were equilibrated for 200 ps at finite temperatures in isothermal–isobaric (NPT) ensemble using MD simulations: interlayer (d) I and (e) II at 300 K and (f) interlayer III at 500 K. The black arrows indicate the atomic displacement relative to the configurations at 0 K. The black dashed circles indicate the empty sites. Ca (large) and Mg (small) atoms are colored brown and green, respectively. Blue and magenta dashed tiles indicate the prismatic stacking fault and perfect lattices, respectively. Prismatic stacking fault structures in C14 Nb(Cr, Ni, Al)\textsubscript{2} Laves phase after (g-h) room temperature impact deformation and (i) high-temperature compression. The white dashed circles indicate the sites with high intensity. (g-i) adapted from \cite{zhang2020shuffle} with permission from the American Physical Society.}
\label{fig2}
\end{figure*}

Three prismatic SF states with different microstructures and energies were found in this work (see Figure \ref{fig2}(a-c)). The structures of SF I and II are mirror symmetric and therefore exhibit the same SFE. The lattice pattern consists of a Laves building block (marked in blue dashed rhombic tile) with the characteristic triple-layer and a Zr\textsubscript{4}Al\textsubscript{3} building block (marked in blue dashed rectangular tile) exists also in $\mu$-phase was found in SF I and II. An empty site in the Zr\textsubscript{4}Al\textsubscript{3} building block was observed, see Figure \ref{fig2}(a,b). The same lattice pattern exists in SF III, however, the site occupancies and chemical distributions are different (see Figure \ref{fig2}(c)). An empty site exists in the middle atomic column of the Laves building block in SF III. Interestingly, these two SF structures are not thermally stable at room or elevated temperature. After equilibrium at 300 K, a structural transition induced by short-range diffusion was observed in SF I and II. The empty site is partially occupied by small (Mg) atoms hopping from a nearby site at the junction of four lattice tiles, and the large (Ca) atoms in the middle atomic column of the Laves building block shift towards the site where the small atoms are left behind, see Figure \ref{fig2}(d-e). The structure of SF I (II) at 300 K is very similar to the experimentally observed prismatic SF structures after room temperature deformation, see Figure \ref{fig2}(g,h). In addition to the identical lattice pattern, the atomic distributions in SFs are similar in our atomic configurations and high-resolution STEM images. 
The positions of three bright atomic columns (highlighted using white dashed circles) corresponding to the heavy elements (Nb in Nb-base Laves phase and Ca in C14 CaMg\textsubscript{2}) correlate well with the simulation results. In addition, the shift of the middle atomic column out of the Laves triple-layer was also reported in the experiment as illustrated using the grey dashed line in Figure \ref{fig2}(g,h).

\begin{figure*}[htbp!]
\centering
\includegraphics[width=\textwidth]{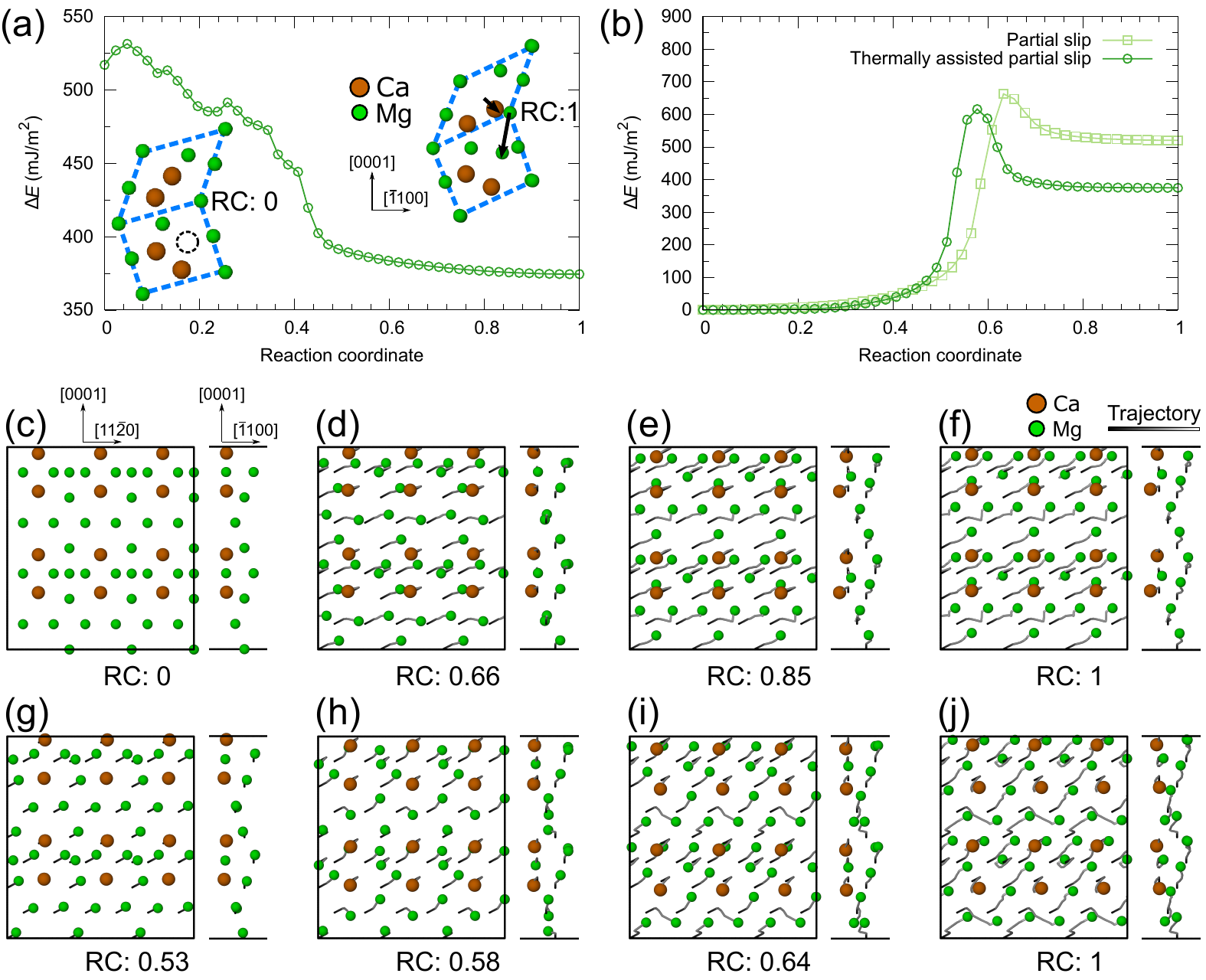}
\caption{Mechanisms of partial slip and formation of stacking fault I (II) at interlayer I (II) in C14 CaMg\textsubscript{2}. (a) Minimum energy path computed for the transition from glide-induced SF I to diffusion-assisted SF I using NEB. The insets show the stacking fault structures at reaction coordinates (RC) 0 and 1. The black arrows indicate the atomic displacement relative to RC 0. The dashed circle indicates the empty site. (b) Minimum energy path computed for the partial slip and thermally assisted partial slip using NEB. (c-f) Transition states of the partial slip. (g-j) Transition states of the thermally assisted partial slip. Left: view along $[\bar{1} 1 0 0]$ direction; right: view along $[1 1 \bar{2} 0]$ direction. Ca (large) and Mg (small) atoms are colored brown and green, respectively. Trajectories of atoms are colored in a black-grey gradient according to the reaction coordinate.}
\label{fig3}
\end{figure*}

According to the atomic displacement introduced by thermalization at room temperature, the structure of glide-induced SF I (II) was optimized by shifting one out of two atoms in the Mg atomic column to the empty site (see the displacement vectors as shown in the inset of Figure \ref{fig3}(a)). After energy minimization, the Ca atoms located in the atomic column in the Laves triple-layer shift towards the empty site left behind by the Mg atoms. The SFE of SF I decreases by 28\% to 374 mJ/m\textsuperscript{2} in CaMg\textsubscript{2}, and by 58\% to 191 mJ/m\textsuperscript{2} in MgZn\textsubscript{2} after the structural opimization. The NEB calculation was performed to find the MEP between the initial and optimized SF states (see Figure \ref{fig3}(a)). For the glide-induced SF I (II), a small energy barrier of 15 mJ/m\textsuperscript{2} needs to be overcome with thermal assistance to reach the diffusion-assisted SF state with lower energy. In addition to a lower SFE, the formation of diffusion-assisted SF I (II) states (Figure \ref{fig3}(b)) is mediated by a different slip mechanism (Figure \ref{fig3}(c-j)) with a lower activation energy. Simple crystallographic slip mediates the partial slip between the pristine and glide-induced SF I (II) states, with all atoms moving in the same direction as the partial Burgers vector (as illustrated in Figure \ref{fig3}(c-f)). For the partial slip between the pristine and diffusion-assisted SF I (II) states, a synchroshear-like mechanism, where certain Mg atoms move in a direction that differs from the direction of the partial Burgers vector, was identified. 

SF III shows better thermal stability than SF I and II, it remains the same atomic structure at 300 K, and a structural transition was observed after equilibrium at 500 K, see Figure \ref{fig2}(c,f). The empty site in the Laves building block is partially occupied by small (Mg) atoms hopping from a nearby site at the junction of lattice tiles. Compared to the SF observed in experiments following high-temperature deformation, diffusion-assisted SF III at 500 K exhibits a similar chemical distribution in the Laves building block but a distinct chemical distribution in the Zr\textsubscript{4}Al\textsubscript{3} building block. In the experimental observation, the heavy (large) atoms are positioned at the up and down sites within the Zr\textsubscript{4}Al\textsubscript{3} building block (Figure \ref{fig2}(i)), whereas in the simulation, they are situated at the left and right sites (Figure \ref{fig2}(f)). The SF structure is optimized by swapping the elemental type in the Zr\textsubscript{4}Al\textsubscript{3} building block to be identical as reported in the experiment (see Figure \ref{figS2}(c,d)). The SFE of the optimized SF III decreases to 326 (by 23 \%) and 189 (by 29 \%) mJ/m\textsuperscript{2} in CaMg\textsubscript{2} and MgZn\textsubscript{2}, respectively.
The formation of the optimized SF III state necessitates a more substantial thermal fluctuation than the optimized SF I (II) state due to its involvement in not only atomic diffusion but also atomic swapping, which rationalizes the experimental observation of optimized SF III mainly after high-temperature deformation \cite{zhang2020shuffle}.

The formation of more energetically favorable prismatic SFs with the assistance of diffusion implies the strong thermally activated nature of prismatic partial dislocations. Thermal assistance was demonstrated to be indispensable in activating synchro-Shockley dislocations, implying that the motion of this zonal dislocation is prohibited at low temperatures \cite{xie2022unveiling,xie2023thermally}. A temperature-dependent deformation behavior is also expected for prismatic partial dislocations where the thermal fluctuation can significantly lower not only the planar fault energies but also the associated energy barriers of slip events.

\begin{figure*}[htbp!]
\centering
\includegraphics[width=\textwidth]{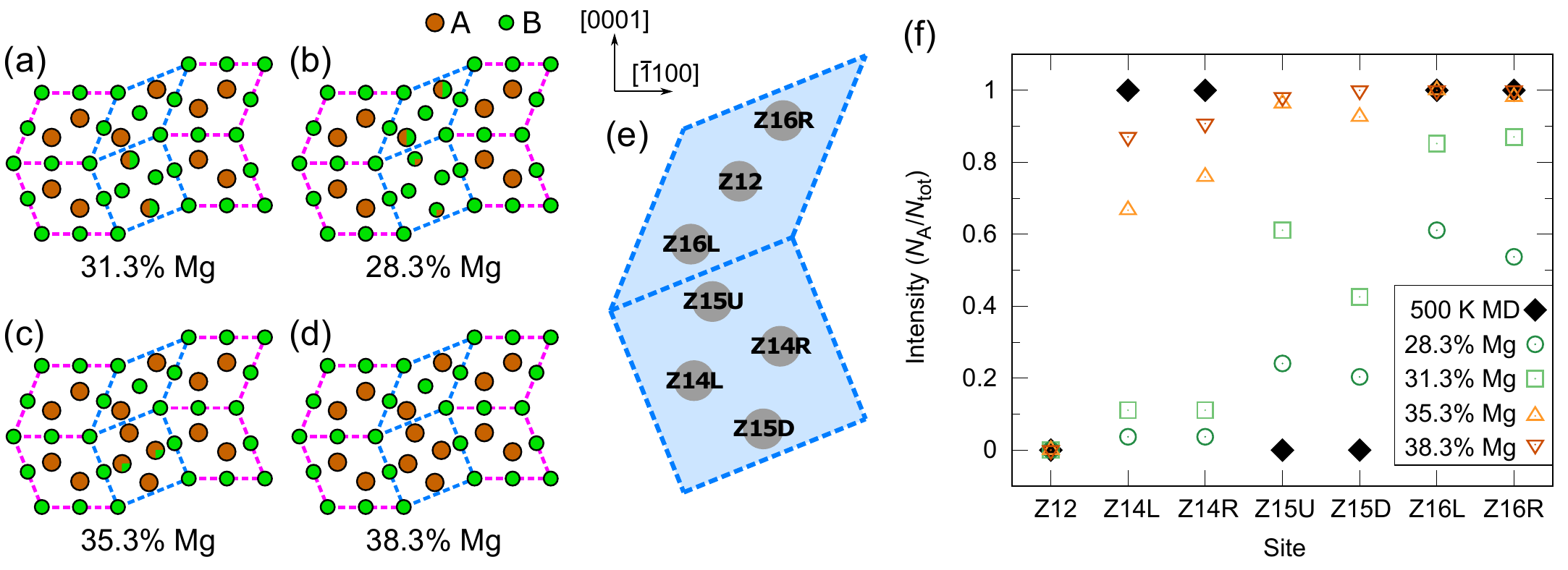}
\caption{Prismatic stacking fault III in C14 MgZn\textsubscript{2} at 500 K and off-stoichiometric compositions. (a-d) Schematic of chemical distribution in the SF III under different target Mg concentrations after 1.6 ns hybrid MD/MC simulations (1 MC step every 0.1 ps MD time) in variance-constrained semi-grand canonical (VC-SGC) ensemble ($T$=500 K, $\kappa$=1000, $\Delta\mu$=$\pm$0.3 eV). Mg (large) and Zn (small) atoms are colored brown and green, respectively. For the atomic sites with two colors, the fraction of the area represents the rough estimation of the percentage of the element in one atomic column. (f) The schematic illustration of the atomic arrangement in prismatic stacking fault. The atomic site is named after the type of surrounding atomic cluster (i.e., Z12 indicates the site is the center of an icosahedral cluster) and corresponding position (i.e.,  L: left; R: right; U: up; D: down) in the tiles. (e) The intensity of A-type atom (for AB\textsubscript{2} Laves phase) in an atomic column of the stacking fault sites. }
\label{fig4}
\end{figure*}

Zhang et al. \cite{zhang2020shuffle} reported the uneven intensities of the atomic columns for the same sites in different unit cells along the prismatic SF plane. This variation can be attributed to the mixture of different SF structures (SF I, II and III) and the inconsistency of the local chemical environment in the same SF structure.
The chemical-dependent prismatic SF III structures in C14 MgZn\textsubscript{2} were investigated using hybrid MD/MC in the variance-constrained semi-grand canonical (VC-SGC) ensemble at 500 K. The target concentrations were set within the range of 5\% off-stoichiometry. After equilibrium, the site occupancy intensity $N\textsubscript{A}/N\textsubscript{tot}$ (where $N\textsubscript{A}$ is the number of A-type atoms in AB\textsubscript{2} Laves phase) was averaged over all unit cells in the prismatic SFs. The statistics of the site occupancies in SF III at different chemical compositions are shown in Figure \ref{fig4}(f) and the naming convention of the sites is illustrated in Figure \ref{fig4}(e).
Multiple SF III states with different chemical compositions are depicted in the chemical distribution schematic of Figure \ref{fig4}(a-d). A significantly higher concentration of solute atoms was identified at prismatic SFs than the matrix after 16,000 MC steps (1.6 ns MD time), see Figure \ref{figS3}. The atomic swapping mainly takes place in the Zr\textsubscript{4}Al\textsubscript{3} building blocks, where the Laves building blocks remain unchanged except at 28.3\% Mg composition. At 31.3\% Mg composition (-2\% off-stoichiometry), Z14L and Z14R are almost occupied by Zn (small) atoms in contrast to the stoichiometric SF III configuration. In addition, half of the Z15U and Z15D sites are Mg (large) atoms. With decreasing Mg content, around 75\% Z15U and Z15D sites are occupied by Zn atoms at 28.3\% Mg composition. On the Mg-rich side of the off-stoichiometry, all sites in the Zr\textsubscript{4}Al\textsubscript{3} building blocks are mostly occupied by Mg atoms. At 38.3\% Mg composition (+5\% off-stoichiometry), nearly pure Mg Zr\textsubscript{4}Al\textsubscript{3} building blocks were obtained. 
The variation of local chemical distribution in SF III at different chemical compositions provides a possible explanation of varying intensity in prismatic SFs as reported in the literature \cite{zhang2020shuffle}. 
The presence of diverse prismatic SF states in off-stoichiometry implies that the chemical composition has an impact on prismatic plasticity. To gain a more profound comprehension of prismatic plasticity, additional works are necessary, such as unraveling dislocation mechanisms and the effects of solute atoms on dislocation motion.

In this study, we evaluated the prismatic slip systems and potential metastable states of C14 CaMg\textsubscript{2} and MgZn\textsubscript{2} using atomistic simulations. Various prismatic stacking fault states with identical lattice patterns but different chemical distributions and site occupancies were identified at varying interlayers, temperatures and chemical compositions. The formation of prismatic stacking faults that are energetically favorable involves short-range diffusion, indicating that prismatic slip is thermally activated. The findings of this work enhance our comprehension of non-basal slip in Laves phases, which is dependent on temperature and chemical composition, and can be generalized to other topologically close-packed phases.

\FloatBarrier

\section*{Acknowledgments}
The authors acknowledge financial support by the Deutsche Forschungsgemeinschaft (DFG) through the projects A02, A05 and C02 of the SFB1394 Structural and Chemical Atomic Complexity – From Defect Phase Diagrams to Material Properties, project ID 409476157. This project has received funding from the European Research Council (ERC) under the European Union’s Horizon 2020 research and innovation programme (grant agreement No. 852096 FunBlocks). Simulations were performed with computing resources granted by RWTH Aachen University under project (p0020267) and by the EXPLOR center of the Université de Lorraine and by the GENCI-TGCC (Grant 2020-A0080911390).
%\bibliography{main}% Produces the bibliography via BibTeX.
\clearpage

\renewcommand{\thefigure}{S\arabic{figure}}
\setcounter{figure}{0}
\setcounter{table}{0}
\renewcommand{\thetable}{S\arabic{table}}
\maketitle

\onecolumngrid
\begin{center}
\textbf{\large Supplementary Material}
\end{center}

\section{Simulation methods}
The atomistic simulations were performed using the molecular dynamics (MD) software package LAMMPS \cite{LAMMPS}. 
The interatomic interactions were modeled by the modified embedded atom method (MEAM) potential by Kim et al. \cite{kim2015modified} for Ca-Mg and the embedded atom method (EAM) potential by Brommer et al. \cite{brommer2009vibrational} for the Mg-Zn system. 
Both potentials reasonably describe the mechanical properties of C14 CaMg\textsubscript{2} and MgZn\textsubscript{2} Laves phases as compared to experiments and $ab$ $initio$ calculations (see Table \ref{tabS.1}).

To investigate the prismatic $\{\bar{1} 1 0 0\}$ slip systems, the C14 CaMg\textsubscript{2} and MgZn\textsubscript{2} Laves structures were constructed using Atomsk \cite{hirel2015atomsk} following the crystallographic orientation: $\mathbf{x}=[ 1 1 \bar{2} 0 ]$, $\mathbf{y}=[ \bar{1} 1 0 0 ]$ and $\mathbf{z}=[ 0 0 0 1 ]$. Generalized stacking fault energy (GSFE) surfaces were calculated by incrementally shifting one-half of the crystal along the slip directions across the slip plane. Periodic boundary conditions (PBC) were applied in the $\mathbf{x}$ and $\mathbf{z}$ directions parallel to the slip plane (contains 3 $\times$ 2 unit cells). Non-periodic boundary conditions were used along the $\mathbf{y}$-$[ \bar{1} 1 0 0 ]$ direction (20 unite cells in $\mathbf{y}$-direction). The FIRE \cite{bitzek2006structural,guenole2020assessment} algorithm with the force tolerance of $10^{-8}$ eV/Å was used to relax atoms along the $\mathbf{y}$-$[ \bar{1} 1 0 0 ]$ direction after each displacement step. The stacking fault states were identified on the GSFE surfaces and then fully relaxed. In addition, climbing image nudged elastic band (NEB) \cite{henkelman2000climbing,henkelman2000improved} calculations were performed to find saddle points and minimum energy paths (MEPs) of slip events. The spring constants for parallel and perpendicular nudging forces are both 1.0 eV/Å$^{2}$. Quickmin \cite{sheppard2008optimization} was used to minimize the energies across all replicas until the force norm was below 0.01 eV/Å.

The prismatic stacking fault configurations (9 $\times$ 40 $\times$ 6 unit cells with PBC in all directions) in C14 CaMg\textsubscript{2} and MgZn\textsubscript{2} were equilibrated for 200 ps at 300 K or 500 K using the Nosé-Hoover thermostat together with the Nosé-Hoover barostat \cite{hoover1985canonical} to relax possible stresses. The timestep is 1 fs. To investigate the chemical distribution at planar faults, hybrid MD/Monte Carlo (MC) simulations in the variance-constrained semi-grand canonical (VC-SGC) \cite{sadigh2012scalable} ensemble ($\kappa$=1000, $\Delta\mu$=$\pm$ 0.3 eV) were performed at $T$=500 K on the C14 MgZn\textsubscript{2} sample contains prismatic stacking faults for 1.6 ns (1 MC step every 0.1 ps MD time). The target concentrations of Mg atoms vary from 28.3\% to 38.3\% ($\pm$ 5\% off-stoichiometry). To ensure that the systems reach equilibrium at the target concentration, chemical potential and temperature, we used the evolution of potential energies (Figure S\ref{figS4}), local and global concentrations (Figure S\ref{figS5}), and successful swap rates (Figure S\ref{figS6}) as indicators.
 
The atomic structures were quenched using the conjugate gradient method to reduce thermal noise for visualization. The Open Visualization Tool (OVITO) \cite{stukowski2009visualization} was used to visualize the atomic configurations and trajectories. 

\begin{table*}[!htbp]
\centering
\caption[]{\label{tabS.1}Potential properties of AB\textsubscript{2} Laves phases using the Ca-Mg MEAM and Mg-Zn EAM potentials. $a\textsubscript{0}$ and $c\textsubscript{0}$: lattice parameters; $C\textsubscript{ij}$: elastic constants; $B$: isotropic bulk modulus (Hill approximation); $G$: isotropic shear modulus (Hill approximation); $\nu$: Poisson's ratio (Hill approximation); $\gamma\textsubscript{SF}$: basal stacking fault energy.}
\centering
\scriptsize
\begin{tabular}{l@{\hspace{1cm}}c@{\hspace{1cm}}c@{\hspace{1cm}}c@{\hspace{1cm}}c@{\hspace{1cm}}c@{\hspace{1cm}}c}
\hline\hline
\addlinespace[0.1cm]
\multicolumn{1}{l}{} & \multicolumn{3}{c}{C14 CaMg\textsubscript{2}} & \multicolumn{3}{c}{C14 MgZn\textsubscript{2}}\\
\cmidrule(lr){2-4} \cmidrule(lr){5-7}
\addlinespace[0.1cm]
Properties &  Experiment & DFT & MEAM & Experiment & DFT & EAM \\
\addlinespace[0.1cm]
\hline
\addlinespace[0.1cm]
$a\textsubscript{0}$ (\text{\AA}) & 6.225 \cite{nowotny1946kristallstrukturen} & 6.230 \cite{arroyave2006intermetallics} & 6.142 & 5.223 \cite{kammerer2015diffusion} & 5.251 \cite{jain2013commentary} & 5.166 \\
$c\textsubscript{0}$ (\text{\AA}) & 10.180 \cite{nowotny1946kristallstrukturen} & 10.070 \cite{arroyave2006intermetallics} & 9.991 & 8.566 \cite{kammerer2015diffusion} & 8.445 \cite{jain2013commentary} & 8.465 \\
$C\textsubscript{11}$ (GPa) & 61.2 \cite{sumer1962elastic} & 59.5 \cite{ganeshan2009elastic} & 70.9 & 107.3 \cite{seidenkranz1976single} & 101 \cite{jain2013commentary} & 120.5 \\
$C\textsubscript{12}$ (GPa) & 17.6 \cite{sumer1962elastic}& 17.8 \cite{ganeshan2009elastic} & 19.6 & 45.5 \cite{seidenkranz1976single} & 52 \cite{jain2013commentary} & 46.1 \\
$C\textsubscript{13}$ (GPa) & 15.0 \cite{sumer1962elastic} & 12.6 \cite{ganeshan2009elastic} & 19.5 & 27.4 \cite{seidenkranz1976single} & 36 \cite{jain2013commentary} & 42.4 \\
$C\textsubscript{33}$ (GPa) & 65.5 \cite{sumer1962elastic} & 66.0 \cite{ganeshan2009elastic} & 70.3 & 126.4 \cite{seidenkranz1976single} & 123 \cite{jain2013commentary} & 118.4 \\
$C\textsubscript{44}$ (GPa) & 19.3 \cite{sumer1962elastic} & 17.4 \cite{ganeshan2009elastic} & 25.5 & 27.7 \cite{seidenkranz1976single} & 25 \cite{jain2013commentary} & 25.2 \\
$C\textsubscript{66}$ (GPa) & 21.8 \cite{sumer1962elastic} & 20.9 \cite{ganeshan2009elastic} & 25.7 & 30.9 \cite{seidenkranz1976single} & 24 \cite{jain2013commentary} & 37.2 \\
$B$ (GPa) & 31.5 \cite{sumer1962elastic} & 30.1 \cite{ganeshan2009elastic} & 36.6 & 60.2 \cite{seidenkranz1976single} & 64 \cite{jain2013commentary} & 69.0 \\
$G$ (GPa) & 21.3 \cite{sumer1962elastic} & 20.4 \cite{ganeshan2009elastic} & 25.6 & 32.6 \cite{seidenkranz1976single} & 27 \cite{jain2013commentary} & 32.1 \\
$\nu$ & 0.22 \cite{sumer1962elastic} & 0.22 \cite{ganeshan2009elastic} & 0.22 & 0.27 \cite{seidenkranz1976single} & 0.31 \cite{jain2013commentary} & 0.30 \\
$\gamma$\textsubscript{SF} (mJ/$\text{m}^{2}$) & - & 21 \cite{Tehranchi}  & 14 & 100 \cite{rennert1977structural} & 104 \cite{ma2013ab} & 50 \\
\hline\hline
\end{tabular}
\end{table*}

\begin{figure*}[htbp!]
\centering
\includegraphics[width=\textwidth]{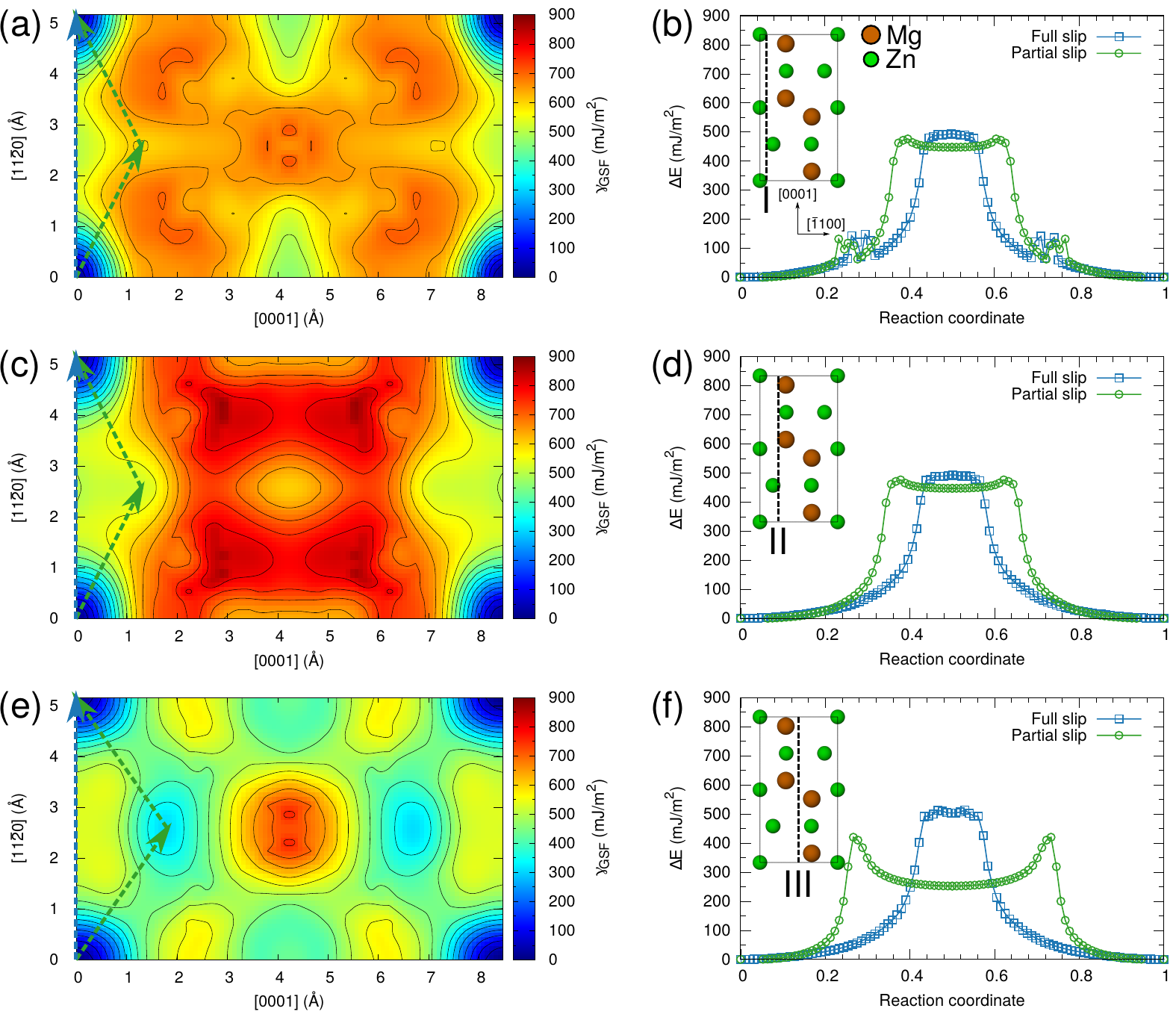}
\caption{Prismatic slip systems of C14 MgZn\textsubscript{2}. Generalized stacking fault energy ($\gamma$) surfaces of interlayer (a) I, (c) II and (e) III. Minimum energy paths computed for full ($[1 1 \bar{2} 0]$ or $\langle a \rangle$) and partial ($\langle a+c \rangle$) slip on interlayer (b) I, (d) II and (f) III using NEB. The paths of full and partial slip are marked in blue and green dashed arrows on the $\gamma$ surfaces, respectively. The C14 unit cell and prismatic slip planes (marked in black dashed line) are shown in the insets. Mg (large) and Zn (small) atoms are colored brown and green, respectively.}
\label{figS1}
\end{figure*}

\begin{figure*}[htbp!]
\centering
\includegraphics[width=0.75\textwidth]{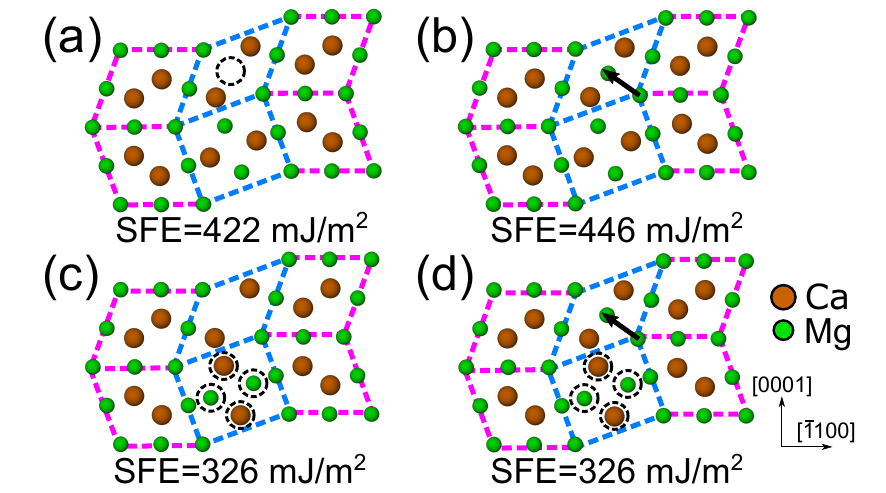}
\caption{Stacking fault III in C14 CaMg\textsubscript{2}. (a) Stacking fault structure after the crystallographic slip. (b) Stacking fault structure after the atomic shuffling of nearby Mg atoms to the Z12 site. (c) Stacking fault structure after the chemical swapping of atoms in the Zr\textsubscript{4}Al\textsubscript{3} building block. (d) Stacking fault structure after the atomic shuffling of nearby Mg atoms to the Z12 site and chemical swapping of atoms in the Zr\textsubscript{4}Al\textsubscript{3} building block. The black arrows indicate the atomic displacement relative to the configuration in (a). The dashed circles indicate (a) the empty sites and (c-d) the sites where chemical swapping is applied. Ca (large) and Mg (small) atoms are colored brown and green, respectively.}
\label{figS2}
\end{figure*}

\begin{figure*}[htbp!]
\centering
\includegraphics[width=0.5\textwidth]{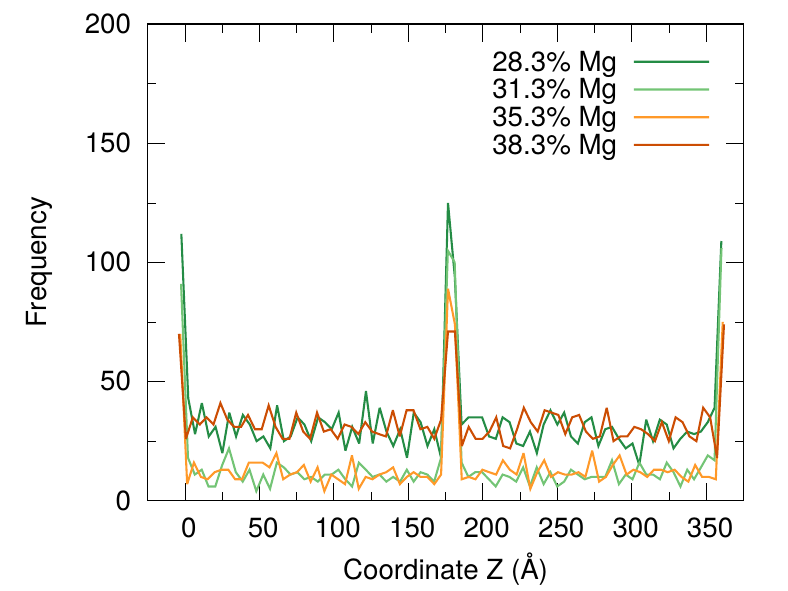}
\caption{Distribution of solute atoms along $[\bar{1} 1 0 0]$ direction (z-axis) in the simulated C14 MgZn\textsubscript{2} after 1600 ps hybrid MD/MC simulations at 500 K.}
\label{figS3}
\end{figure*}

\begin{figure*}[htbp!]
\centering
\includegraphics[width=0.5\textwidth]{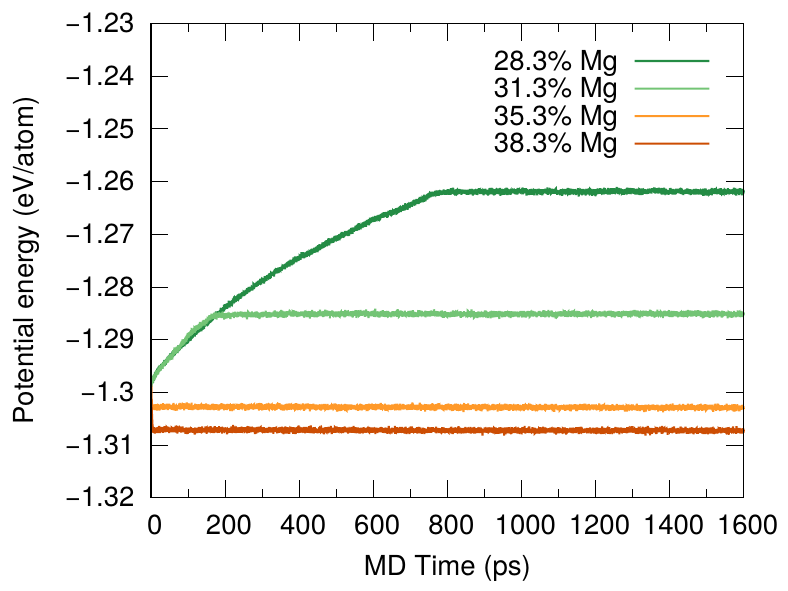}
\caption{Evolution of potential energy of the simulated C14 MgZn\textsubscript{2} during the hybrid MD/MC simulations at 500 K.}
\label{figS4}
\end{figure*}

\begin{figure*}[htbp!]
\centering
\includegraphics[width=\textwidth]{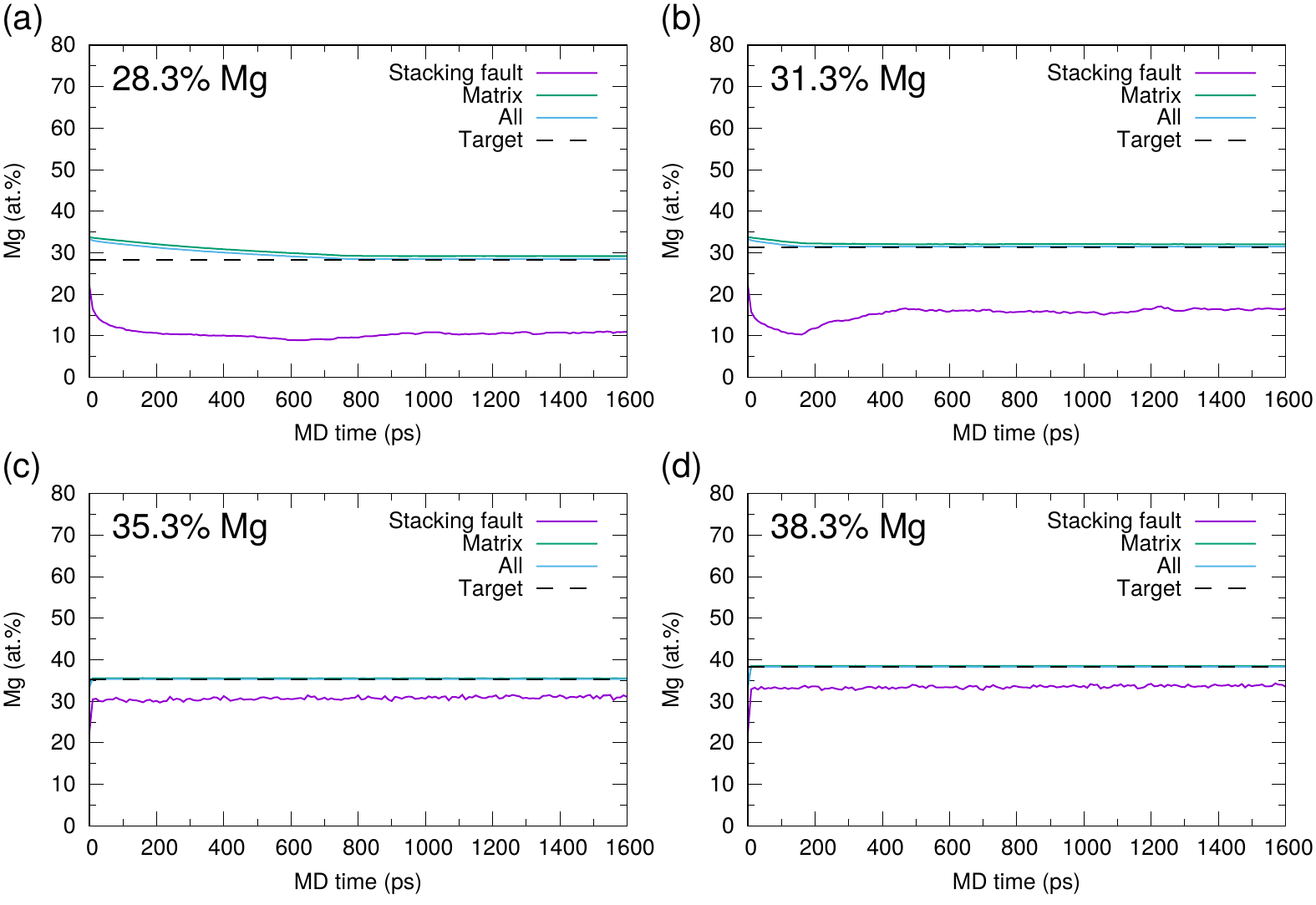}
\caption{Evolution of local and global Mg concentrations of the simulated C14 MgZn\textsubscript{2} during the hybrid MD/MC simulations at 500 K.}
\label{figS5}
\end{figure*}

\begin{figure*}[htbp!]
\centering
\includegraphics[width=\textwidth]{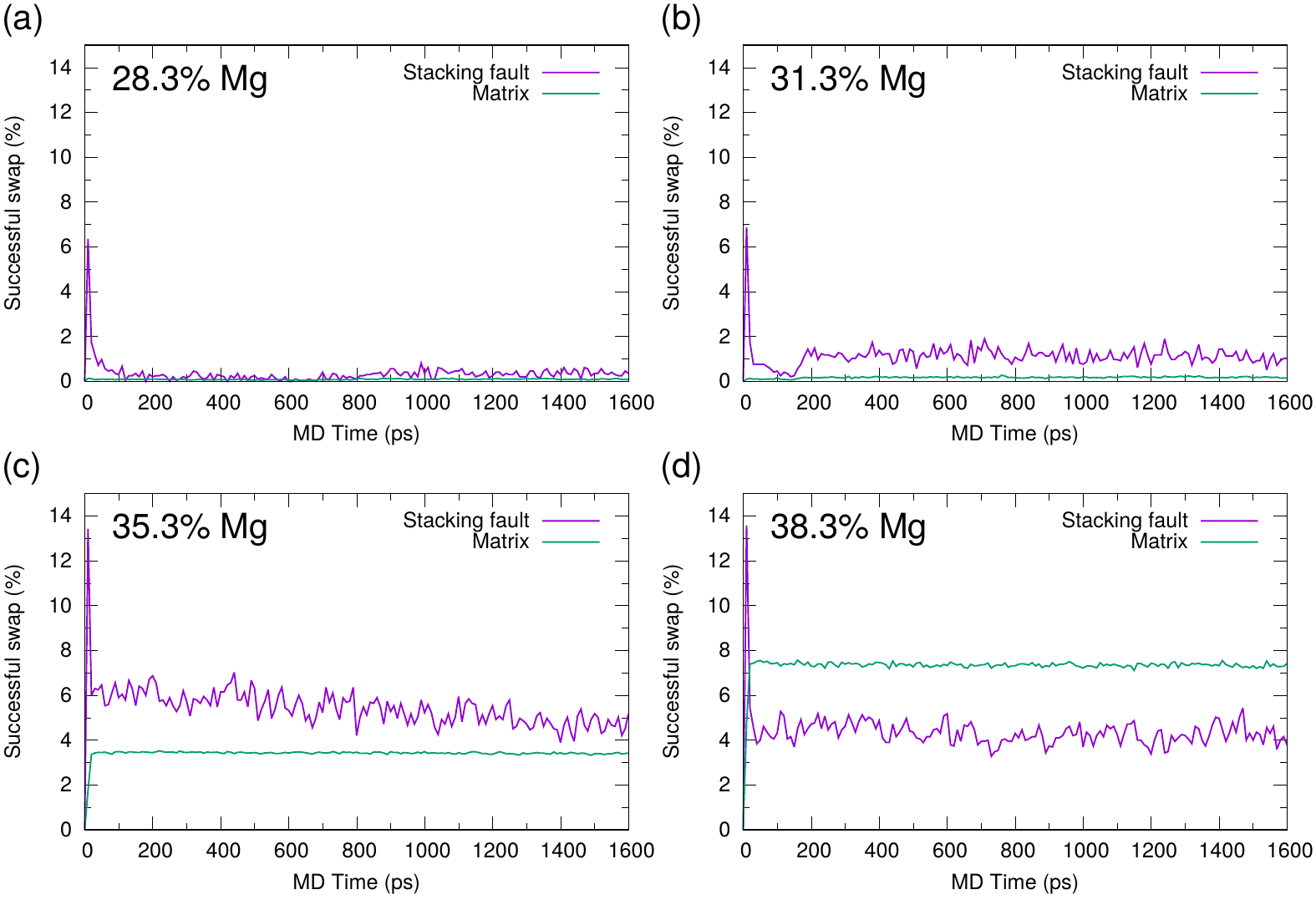}
\caption{Evolution of successful swap of the simulated C14 MgZn\textsubscript{2} in every 10 ps during the hybrid MD/MC simulations at 500 K.}
\label{figS6}
\end{figure*}

\clearpage

\bibliography{main}

%apsrev4-2.bst 2019-01-14 (MD) hand-edited version of apsrev4-1.bst
%Control: key (0)
%Control: author (8) initials jnrlst
%Control: editor formatted (1) identically to author
%Control: production of article title (0) allowed
%Control: page (0) single
%Control: year (1) truncated
%Control: production of eprint (0) enabled
\begin{thebibliography}{38}%
\makeatletter
\providecommand \@ifxundefined [1]{%
 \@ifx{#1\undefined}
}%
\providecommand \@ifnum [1]{%
 \ifnum #1\expandafter \@firstoftwo
 \else \expandafter \@secondoftwo
 \fi
}%
\providecommand \@ifx [1]{%
 \ifx #1\expandafter \@firstoftwo
 \else \expandafter \@secondoftwo
 \fi
}%
\providecommand \natexlab [1]{#1}%
\providecommand \enquote  [1]{``#1''}%
\providecommand \bibnamefont  [1]{#1}%
\providecommand \bibfnamefont [1]{#1}%
\providecommand \citenamefont [1]{#1}%
\providecommand \href@noop [0]{\@secondoftwo}%
\providecommand \href [0]{\begingroup \@sanitize@url \@href}%
\providecommand \@href[1]{\@@startlink{#1}\@@href}%
\providecommand \@@href[1]{\endgroup#1\@@endlink}%
\providecommand \@sanitize@url [0]{\catcode `\\12\catcode `\$12\catcode
  `\&12\catcode `\#12\catcode `\^12\catcode `\_12\catcode `\%12\relax}%
\providecommand \@@startlink[1]{}%
\providecommand \@@endlink[0]{}%
\providecommand \url  [0]{\begingroup\@sanitize@url \@url }%
\providecommand \@url [1]{\endgroup\@href {#1}{\urlprefix }}%
\providecommand \urlprefix  [0]{URL }%
\providecommand \Eprint [0]{\href }%
\providecommand \doibase [0]{https://doi.org/}%
\providecommand \selectlanguage [0]{\@gobble}%
\providecommand \bibinfo  [0]{\@secondoftwo}%
\providecommand \bibfield  [0]{\@secondoftwo}%
\providecommand \translation [1]{[#1]}%
\providecommand \BibitemOpen [0]{}%
\providecommand \bibitemStop [0]{}%
\providecommand \bibitemNoStop [0]{.\EOS\space}%
\providecommand \EOS [0]{\spacefactor3000\relax}%
\providecommand \BibitemShut  [1]{\csname bibitem#1\endcsname}%
\let\auto@bib@innerbib\@empty
%</preamble>
\bibitem [{\citenamefont {Sinha}(1972)}]{sinha1972topologically}%
  \BibitemOpen
  \bibfield  {author} {\bibinfo {author} {\bibfnamefont {A.~K.}\ \bibnamefont
  {Sinha}},\ }\bibfield  {title} {\bibinfo {title} {Topologically close-packed
  structures of transition metal alloys},\ }\href@noop {} {\bibfield  {journal}
  {\bibinfo  {journal} {Progress in Materials Science}\ }\textbf {\bibinfo
  {volume} {15}},\ \bibinfo {pages} {81} (\bibinfo {year} {1972})}\BibitemShut
  {NoStop}%
\bibitem [{\citenamefont {Paufler}(2011)}]{paufler2011early}%
  \BibitemOpen
  \bibfield  {author} {\bibinfo {author} {\bibfnamefont {P.}~\bibnamefont
  {Paufler}},\ }\bibfield  {title} {\bibinfo {title} {Early work on {L}aves
  phases in {E}ast {G}ermany},\ }\href@noop {} {\bibfield  {journal} {\bibinfo
  {journal} {Intermetallics}\ }\textbf {\bibinfo {volume} {19}},\ \bibinfo
  {pages} {599} (\bibinfo {year} {2011})}\BibitemShut {NoStop}%
\bibitem [{\citenamefont {Stein}\ and\ \citenamefont
  {Leineweber}(2021)}]{stein2021laves}%
  \BibitemOpen
  \bibfield  {author} {\bibinfo {author} {\bibfnamefont {F.}~\bibnamefont
  {Stein}}\ and\ \bibinfo {author} {\bibfnamefont {A.}~\bibnamefont
  {Leineweber}},\ }\bibfield  {title} {\bibinfo {title} {Laves phases: a review
  of their functional and structural applications and an improved fundamental
  understanding of stability and properties},\ }\href@noop {} {\bibfield
  {journal} {\bibinfo  {journal} {Journal of Materials Science}\ }\textbf
  {\bibinfo {volume} {56}},\ \bibinfo {pages} {5321} (\bibinfo {year}
  {2021})}\BibitemShut {NoStop}%
\bibitem [{\citenamefont {Livingston}(1992)}]{livingston1992laves}%
  \BibitemOpen
  \bibfield  {author} {\bibinfo {author} {\bibfnamefont {J.}~\bibnamefont
  {Livingston}},\ }\bibfield  {title} {\bibinfo {title} {Laves-phase
  superalloys?},\ }\href@noop {} {\bibfield  {journal} {\bibinfo  {journal}
  {Physica Status Solidi A}\ }\textbf {\bibinfo {volume} {131}},\ \bibinfo
  {pages} {415} (\bibinfo {year} {1992})}\BibitemShut {NoStop}%
\bibitem [{\citenamefont {Pollock}(2010)}]{pollock2010weight}%
  \BibitemOpen
  \bibfield  {author} {\bibinfo {author} {\bibfnamefont {T.~M.}\ \bibnamefont
  {Pollock}},\ }\bibfield  {title} {\bibinfo {title} {Weight loss with
  magnesium alloys},\ }\href@noop {} {\bibfield  {journal} {\bibinfo  {journal}
  {Science}\ }\textbf {\bibinfo {volume} {328}},\ \bibinfo {pages} {986}
  (\bibinfo {year} {2010})}\BibitemShut {NoStop}%
\bibitem [{\citenamefont {Kronberg}(1957)}]{kronberg1957plastic}%
  \BibitemOpen
  \bibfield  {author} {\bibinfo {author} {\bibfnamefont {M.}~\bibnamefont
  {Kronberg}},\ }\bibfield  {title} {\bibinfo {title} {Plastic deformation of
  single crystals of sapphire: basal slip and twinning},\ }\href@noop {}
  {\bibfield  {journal} {\bibinfo  {journal} {Acta Metallurgica}\ }\textbf
  {\bibinfo {volume} {5}},\ \bibinfo {pages} {507} (\bibinfo {year}
  {1957})}\BibitemShut {NoStop}%
\bibitem [{\citenamefont {Chisholm}\ \emph {et~al.}(2005)\citenamefont
  {Chisholm}, \citenamefont {Kumar},\ and\ \citenamefont
  {Hazzledine}}]{chisholm2005dislocations}%
  \BibitemOpen
  \bibfield  {author} {\bibinfo {author} {\bibfnamefont {M.~F.}\ \bibnamefont
  {Chisholm}}, \bibinfo {author} {\bibfnamefont {S.}~\bibnamefont {Kumar}},\
  and\ \bibinfo {author} {\bibfnamefont {P.}~\bibnamefont {Hazzledine}},\
  }\bibfield  {title} {\bibinfo {title} {Dislocations in complex materials},\
  }\href@noop {} {\bibfield  {journal} {\bibinfo  {journal} {Science}\ }\textbf
  {\bibinfo {volume} {307}},\ \bibinfo {pages} {701} (\bibinfo {year}
  {2005})}\BibitemShut {NoStop}%
\bibitem [{\citenamefont {Vedmedenko}\ \emph {et~al.}(2008)\citenamefont
  {Vedmedenko}, \citenamefont {R{\"o}sch},\ and\ \citenamefont
  {Els{\"a}sser}}]{vedmedenko2008first}%
  \BibitemOpen
  \bibfield  {author} {\bibinfo {author} {\bibfnamefont {O.}~\bibnamefont
  {Vedmedenko}}, \bibinfo {author} {\bibfnamefont {F.}~\bibnamefont
  {R{\"o}sch}},\ and\ \bibinfo {author} {\bibfnamefont {C.}~\bibnamefont
  {Els{\"a}sser}},\ }\bibfield  {title} {\bibinfo {title} {First-principles
  density functional theory study of phase transformations in {N}b{C}r2 and
  {T}a{C}r2},\ }\href@noop {} {\bibfield  {journal} {\bibinfo  {journal} {Acta
  materialia}\ }\textbf {\bibinfo {volume} {56}},\ \bibinfo {pages} {4984}
  (\bibinfo {year} {2008})}\BibitemShut {NoStop}%
\bibitem [{\citenamefont {Gu{\'e}nol{\'e}}\ \emph {et~al.}(2019)\citenamefont
  {Gu{\'e}nol{\'e}}, \citenamefont {Mouhib}, \citenamefont {Huber},
  \citenamefont {Grabowski},\ and\ \citenamefont
  {Korte-Kerzel}}]{guenole2019basal}%
  \BibitemOpen
  \bibfield  {author} {\bibinfo {author} {\bibfnamefont {J.}~\bibnamefont
  {Gu{\'e}nol{\'e}}}, \bibinfo {author} {\bibfnamefont {F.-Z.}\ \bibnamefont
  {Mouhib}}, \bibinfo {author} {\bibfnamefont {L.}~\bibnamefont {Huber}},
  \bibinfo {author} {\bibfnamefont {B.}~\bibnamefont {Grabowski}},\ and\
  \bibinfo {author} {\bibfnamefont {S.}~\bibnamefont {Korte-Kerzel}},\
  }\bibfield  {title} {\bibinfo {title} {Basal slip in {L}aves phases: the
  synchroshear dislocation},\ }\href@noop {} {\bibfield  {journal} {\bibinfo
  {journal} {Scripta Materialia}\ }\textbf {\bibinfo {volume} {166}},\ \bibinfo
  {pages} {134} (\bibinfo {year} {2019})}\BibitemShut {NoStop}%
\bibitem [{\citenamefont {Xie}\ \emph {et~al.}(2022)\citenamefont {Xie},
  \citenamefont {Chauraud}, \citenamefont {Atila}, \citenamefont {Bitzek},
  \citenamefont {Korte-Kerzel},\ and\ \citenamefont
  {Gu{\'e}nol{\'e}}}]{xie2022unveiling}%
  \BibitemOpen
  \bibfield  {author} {\bibinfo {author} {\bibfnamefont {Z.}~\bibnamefont
  {Xie}}, \bibinfo {author} {\bibfnamefont {D.}~\bibnamefont {Chauraud}},
  \bibinfo {author} {\bibfnamefont {A.}~\bibnamefont {Atila}}, \bibinfo
  {author} {\bibfnamefont {E.}~\bibnamefont {Bitzek}}, \bibinfo {author}
  {\bibfnamefont {S.}~\bibnamefont {Korte-Kerzel}},\ and\ \bibinfo {author}
  {\bibfnamefont {J.}~\bibnamefont {Gu{\'e}nol{\'e}}},\ }\bibfield  {title}
  {\bibinfo {title} {Unveiling the mechanisms of motion of synchro-shockley
  dislocations in {L}aves phases},\ }\href@noop {} {\bibfield  {journal}
  {\bibinfo  {journal} {arXiv preprint arXiv:2205.02669}\ } (\bibinfo {year}
  {2022})}\BibitemShut {NoStop}%
\bibitem [{\citenamefont {Xie}\ \emph {et~al.}(2023)\citenamefont {Xie},
  \citenamefont {Chauraud}, \citenamefont {Atila}, \citenamefont {Bitzek},
  \citenamefont {Korte-Kerzel},\ and\ \citenamefont
  {Gu{\'e}nol{\'e}}}]{xie2023thermally}%
  \BibitemOpen
  \bibfield  {author} {\bibinfo {author} {\bibfnamefont {Z.}~\bibnamefont
  {Xie}}, \bibinfo {author} {\bibfnamefont {D.}~\bibnamefont {Chauraud}},
  \bibinfo {author} {\bibfnamefont {A.}~\bibnamefont {Atila}}, \bibinfo
  {author} {\bibfnamefont {E.}~\bibnamefont {Bitzek}}, \bibinfo {author}
  {\bibfnamefont {S.}~\bibnamefont {Korte-Kerzel}},\ and\ \bibinfo {author}
  {\bibfnamefont {J.}~\bibnamefont {Gu{\'e}nol{\'e}}},\ }\bibfield  {title}
  {\bibinfo {title} {Thermally activated nature of synchro-{S}hockley
  dislocations in {L}aves phases},\ }\href@noop {} {\bibfield  {journal}
  {\bibinfo  {journal} {arXiv preprint arXiv:2303.03044}\ } (\bibinfo {year}
  {2023})}\BibitemShut {NoStop}%
\bibitem [{\citenamefont {Paufler}\ \emph {et~al.}(1970)\citenamefont
  {Paufler}, \citenamefont {Marschner},\ and\ \citenamefont
  {Schulze}}]{paufler1970mobility}%
  \BibitemOpen
  \bibfield  {author} {\bibinfo {author} {\bibfnamefont {P.}~\bibnamefont
  {Paufler}}, \bibinfo {author} {\bibfnamefont {J.}~\bibnamefont {Marschner}},\
  and\ \bibinfo {author} {\bibfnamefont {G.}~\bibnamefont {Schulze}},\
  }\bibfield  {title} {\bibinfo {title} {The mobility of grown-in dislocations
  in the intermetallic compound {M}g{Z}n2 {I}. stress dependence for edge
  dislocations in prism slip at 390° {C}},\ }\href@noop {} {\bibfield
  {journal} {\bibinfo  {journal} {Physica Status Solidi (b)}\ }\textbf
  {\bibinfo {volume} {40}},\ \bibinfo {pages} {573} (\bibinfo {year}
  {1970})}\BibitemShut {NoStop}%
\bibitem [{\citenamefont {Kubsch}\ \emph {et~al.}(1974)\citenamefont {Kubsch},
  \citenamefont {Paufler},\ and\ \citenamefont {Schulze}}]{kubsch1974mobility}%
  \BibitemOpen
  \bibfield  {author} {\bibinfo {author} {\bibfnamefont {H.}~\bibnamefont
  {Kubsch}}, \bibinfo {author} {\bibfnamefont {P.}~\bibnamefont {Paufler}},\
  and\ \bibinfo {author} {\bibfnamefont {G.}~\bibnamefont {Schulze}},\
  }\bibfield  {title} {\bibinfo {title} {The mobility of grown-in dislocations
  in the intermetallic compound {M}g{Z}n2 during prismatic slip},\ }\href@noop
  {} {\bibfield  {journal} {\bibinfo  {journal} {Physica Status Solidi (a)}\
  }\textbf {\bibinfo {volume} {25}},\ \bibinfo {pages} {269} (\bibinfo {year}
  {1974})}\BibitemShut {NoStop}%
\bibitem [{\citenamefont {Zehnder}\ \emph {et~al.}(2019)\citenamefont
  {Zehnder}, \citenamefont {Czerwinski}, \citenamefont {Molodov}, \citenamefont
  {Sandl{\"o}bes-Haut}, \citenamefont {Gibson},\ and\ \citenamefont
  {Korte-Kerzel}}]{zehnder2019plastic}%
  \BibitemOpen
  \bibfield  {author} {\bibinfo {author} {\bibfnamefont {C.}~\bibnamefont
  {Zehnder}}, \bibinfo {author} {\bibfnamefont {K.}~\bibnamefont {Czerwinski}},
  \bibinfo {author} {\bibfnamefont {K.~D.}\ \bibnamefont {Molodov}}, \bibinfo
  {author} {\bibfnamefont {S.}~\bibnamefont {Sandl{\"o}bes-Haut}}, \bibinfo
  {author} {\bibfnamefont {J.~S.-L.}\ \bibnamefont {Gibson}},\ and\ \bibinfo
  {author} {\bibfnamefont {S.}~\bibnamefont {Korte-Kerzel}},\ }\bibfield
  {title} {\bibinfo {title} {Plastic deformation of single crystalline {C}14
  {M}g2{C}a laves phase at room temperature},\ }\href@noop {} {\bibfield
  {journal} {\bibinfo  {journal} {Materials Science and Engineering: A}\
  }\textbf {\bibinfo {volume} {759}},\ \bibinfo {pages} {754} (\bibinfo {year}
  {2019})}\BibitemShut {NoStop}%
\bibitem [{\citenamefont {Freund}\ \emph {et~al.}(2021)\citenamefont {Freund},
  \citenamefont {Andre}, \citenamefont {Zehnder}, \citenamefont {Rempel},
  \citenamefont {Gerber}, \citenamefont {Zubair}, \citenamefont
  {Sandl{\"o}bes-Haut}, \citenamefont {Gibson},\ and\ \citenamefont
  {Korte-Kerzel}}]{freund2021plastic}%
  \BibitemOpen
  \bibfield  {author} {\bibinfo {author} {\bibfnamefont {M.}~\bibnamefont
  {Freund}}, \bibinfo {author} {\bibfnamefont {D.}~\bibnamefont {Andre}},
  \bibinfo {author} {\bibfnamefont {C.}~\bibnamefont {Zehnder}}, \bibinfo
  {author} {\bibfnamefont {H.}~\bibnamefont {Rempel}}, \bibinfo {author}
  {\bibfnamefont {D.}~\bibnamefont {Gerber}}, \bibinfo {author} {\bibfnamefont
  {M.}~\bibnamefont {Zubair}}, \bibinfo {author} {\bibfnamefont
  {S.}~\bibnamefont {Sandl{\"o}bes-Haut}}, \bibinfo {author} {\bibfnamefont
  {J.~S.-L.}\ \bibnamefont {Gibson}},\ and\ \bibinfo {author} {\bibfnamefont
  {S.}~\bibnamefont {Korte-Kerzel}},\ }\bibfield  {title} {\bibinfo {title}
  {Plastic deformation of the {C}a{M}g2 {C}14-{L}aves phase from 50-250° c},\
  }\href@noop {} {\bibfield  {journal} {\bibinfo  {journal} {Materialia}\
  }\textbf {\bibinfo {volume} {20}},\ \bibinfo {pages} {101237} (\bibinfo
  {year} {2021})}\BibitemShut {NoStop}%
\bibitem [{\citenamefont {Zhang}\ \emph {et~al.}(2020)\citenamefont {Zhang},
  \citenamefont {Zhang}, \citenamefont {Du}, \citenamefont {Li}, \citenamefont
  {Sheng}, \citenamefont {Ye},\ and\ \citenamefont {Du}}]{zhang2020shuffle}%
  \BibitemOpen
  \bibfield  {author} {\bibinfo {author} {\bibfnamefont {Y.}~\bibnamefont
  {Zhang}}, \bibinfo {author} {\bibfnamefont {W.}~\bibnamefont {Zhang}},
  \bibinfo {author} {\bibfnamefont {B.}~\bibnamefont {Du}}, \bibinfo {author}
  {\bibfnamefont {W.}~\bibnamefont {Li}}, \bibinfo {author} {\bibfnamefont
  {L.}~\bibnamefont {Sheng}}, \bibinfo {author} {\bibfnamefont
  {H.}~\bibnamefont {Ye}},\ and\ \bibinfo {author} {\bibfnamefont
  {K.}~\bibnamefont {Du}},\ }\bibfield  {title} {\bibinfo {title} {Shuffle and
  glide mechanisms of prismatic dislocations in a hexagonal {C}14-type
  {L}aves-phase intermetallic compound},\ }\href@noop {} {\bibfield  {journal}
  {\bibinfo  {journal} {Physical Review B}\ }\textbf {\bibinfo {volume}
  {102}},\ \bibinfo {pages} {134117} (\bibinfo {year} {2020})}\BibitemShut
  {NoStop}%
\bibitem [{\citenamefont {Kim}\ \emph {et~al.}(2015)\citenamefont {Kim},
  \citenamefont {Jeon},\ and\ \citenamefont {Lee}}]{kim2015modified}%
  \BibitemOpen
  \bibfield  {author} {\bibinfo {author} {\bibfnamefont {K.-H.}\ \bibnamefont
  {Kim}}, \bibinfo {author} {\bibfnamefont {J.~B.}\ \bibnamefont {Jeon}},\ and\
  \bibinfo {author} {\bibfnamefont {B.-J.}\ \bibnamefont {Lee}},\ }\bibfield
  {title} {\bibinfo {title} {Modified embedded-atom method interatomic
  potentials for {M}g--{X} ({X}={Y}, {S}n, {C}a) binary systems},\ }\href@noop
  {} {\bibfield  {journal} {\bibinfo  {journal} {Calphad}\ }\textbf {\bibinfo
  {volume} {48}},\ \bibinfo {pages} {27} (\bibinfo {year} {2015})}\BibitemShut
  {NoStop}%
\bibitem [{\citenamefont {Brommer}\ \emph {et~al.}(2009)\citenamefont
  {Brommer}, \citenamefont {Boissieu}, \citenamefont {Euchner}, \citenamefont
  {Francoual}, \citenamefont {G{\"a}hler}, \citenamefont {Johnson},
  \citenamefont {Parlinski},\ and\ \citenamefont
  {Schmalzl}}]{brommer2009vibrational}%
  \BibitemOpen
  \bibfield  {author} {\bibinfo {author} {\bibfnamefont {P.}~\bibnamefont
  {Brommer}}, \bibinfo {author} {\bibfnamefont {M.~d.}\ \bibnamefont
  {Boissieu}}, \bibinfo {author} {\bibfnamefont {H.}~\bibnamefont {Euchner}},
  \bibinfo {author} {\bibfnamefont {S.}~\bibnamefont {Francoual}}, \bibinfo
  {author} {\bibfnamefont {F.}~\bibnamefont {G{\"a}hler}}, \bibinfo {author}
  {\bibfnamefont {M.}~\bibnamefont {Johnson}}, \bibinfo {author} {\bibfnamefont
  {K.}~\bibnamefont {Parlinski}},\ and\ \bibinfo {author} {\bibfnamefont
  {K.}~\bibnamefont {Schmalzl}},\ }\bibfield  {title} {\bibinfo {title}
  {Vibrational properties of {M}g{Z}n2},\ }\href@noop {} {\bibfield  {journal}
  {\bibinfo  {journal} {Zeitschrift f{\"u}r Kristallographie-Crystalline
  Materials}\ }\textbf {\bibinfo {volume} {224}},\ \bibinfo {pages} {97}
  (\bibinfo {year} {2009})}\BibitemShut {NoStop}%
\bibitem [{\citenamefont {Henkelman}\ \emph {et~al.}(2000)\citenamefont
  {Henkelman}, \citenamefont {Uberuaga},\ and\ \citenamefont
  {J{\'o}nsson}}]{henkelman2000climbing}%
  \BibitemOpen
  \bibfield  {author} {\bibinfo {author} {\bibfnamefont {G.}~\bibnamefont
  {Henkelman}}, \bibinfo {author} {\bibfnamefont {B.~P.}\ \bibnamefont
  {Uberuaga}},\ and\ \bibinfo {author} {\bibfnamefont {H.}~\bibnamefont
  {J{\'o}nsson}},\ }\bibfield  {title} {\bibinfo {title} {A climbing image
  nudged elastic band method for finding saddle points and minimum energy
  paths},\ }\href@noop {} {\bibfield  {journal} {\bibinfo  {journal} {The
  Journal of chemical physics}\ }\textbf {\bibinfo {volume} {113}},\ \bibinfo
  {pages} {9901} (\bibinfo {year} {2000})}\BibitemShut {NoStop}%
\bibitem [{\citenamefont {Henkelman}\ and\ \citenamefont
  {J{\'o}nsson}(2000)}]{henkelman2000improved}%
  \BibitemOpen
  \bibfield  {author} {\bibinfo {author} {\bibfnamefont {G.}~\bibnamefont
  {Henkelman}}\ and\ \bibinfo {author} {\bibfnamefont {H.}~\bibnamefont
  {J{\'o}nsson}},\ }\bibfield  {title} {\bibinfo {title} {Improved tangent
  estimate in the nudged elastic band method for finding minimum energy paths
  and saddle points},\ }\href@noop {} {\bibfield  {journal} {\bibinfo
  {journal} {The Journal of chemical physics}\ }\textbf {\bibinfo {volume}
  {113}},\ \bibinfo {pages} {9978} (\bibinfo {year} {2000})}\BibitemShut
  {NoStop}%
\bibitem [{\citenamefont {Bitzek}\ \emph {et~al.}(2006)\citenamefont {Bitzek},
  \citenamefont {Koskinen}, \citenamefont {G{\"a}hler}, \citenamefont
  {Moseler},\ and\ \citenamefont {Gumbsch}}]{bitzek2006structural}%
  \BibitemOpen
  \bibfield  {author} {\bibinfo {author} {\bibfnamefont {E.}~\bibnamefont
  {Bitzek}}, \bibinfo {author} {\bibfnamefont {P.}~\bibnamefont {Koskinen}},
  \bibinfo {author} {\bibfnamefont {F.}~\bibnamefont {G{\"a}hler}}, \bibinfo
  {author} {\bibfnamefont {M.}~\bibnamefont {Moseler}},\ and\ \bibinfo {author}
  {\bibfnamefont {P.}~\bibnamefont {Gumbsch}},\ }\bibfield  {title} {\bibinfo
  {title} {Structural relaxation made simple},\ }\href@noop {} {\bibfield
  {journal} {\bibinfo  {journal} {Physical review letters}\ }\textbf {\bibinfo
  {volume} {97}},\ \bibinfo {pages} {170201} (\bibinfo {year}
  {2006})}\BibitemShut {NoStop}%
\bibitem [{\citenamefont {Gu{\'e}nol{\'e}}\ \emph {et~al.}(2020)\citenamefont
  {Gu{\'e}nol{\'e}}, \citenamefont {N{\"o}hring}, \citenamefont {Vaid},
  \citenamefont {Houll{\'e}}, \citenamefont {Xie}, \citenamefont {Prakash},\
  and\ \citenamefont {Bitzek}}]{guenole2020assessment}%
  \BibitemOpen
  \bibfield  {author} {\bibinfo {author} {\bibfnamefont {J.}~\bibnamefont
  {Gu{\'e}nol{\'e}}}, \bibinfo {author} {\bibfnamefont {W.~G.}\ \bibnamefont
  {N{\"o}hring}}, \bibinfo {author} {\bibfnamefont {A.}~\bibnamefont {Vaid}},
  \bibinfo {author} {\bibfnamefont {F.}~\bibnamefont {Houll{\'e}}}, \bibinfo
  {author} {\bibfnamefont {Z.}~\bibnamefont {Xie}}, \bibinfo {author}
  {\bibfnamefont {A.}~\bibnamefont {Prakash}},\ and\ \bibinfo {author}
  {\bibfnamefont {E.}~\bibnamefont {Bitzek}},\ }\bibfield  {title} {\bibinfo
  {title} {Assessment and optimization of the fast inertial relaxation engine
  (fire) for energy minimization in atomistic simulations and its
  implementation in lammps},\ }\href@noop {} {\bibfield  {journal} {\bibinfo
  {journal} {Computational Materials Science}\ }\textbf {\bibinfo {volume}
  {175}},\ \bibinfo {pages} {109584} (\bibinfo {year} {2020})}\BibitemShut
  {NoStop}%
\bibitem [{\citenamefont {Sheppard}\ \emph {et~al.}(2008)\citenamefont
  {Sheppard}, \citenamefont {Terrell},\ and\ \citenamefont
  {Henkelman}}]{sheppard2008optimization}%
  \BibitemOpen
  \bibfield  {author} {\bibinfo {author} {\bibfnamefont {D.}~\bibnamefont
  {Sheppard}}, \bibinfo {author} {\bibfnamefont {R.}~\bibnamefont {Terrell}},\
  and\ \bibinfo {author} {\bibfnamefont {G.}~\bibnamefont {Henkelman}},\
  }\bibfield  {title} {\bibinfo {title} {Optimization methods for finding
  minimum energy paths},\ }\href@noop {} {\bibfield  {journal} {\bibinfo
  {journal} {The Journal of chemical physics}\ }\textbf {\bibinfo {volume}
  {128}},\ \bibinfo {pages} {134106} (\bibinfo {year} {2008})}\BibitemShut
  {NoStop}%
\bibitem [{\citenamefont {Thompson}\ \emph {et~al.}(2022)\citenamefont
  {Thompson}, \citenamefont {Aktulga}, \citenamefont {Berger}, \citenamefont
  {Bolintineanu}, \citenamefont {Brown}, \citenamefont {Crozier}, \citenamefont
  {in~'t Veld}, \citenamefont {Kohlmeyer}, \citenamefont {Moore}, \citenamefont
  {Nguyen}, \citenamefont {Shan}, \citenamefont {Stevens}, \citenamefont
  {Tranchida}, \citenamefont {Trott},\ and\ \citenamefont {Plimpton}}]{LAMMPS}%
  \BibitemOpen
  \bibfield  {author} {\bibinfo {author} {\bibfnamefont {A.~P.}\ \bibnamefont
  {Thompson}}, \bibinfo {author} {\bibfnamefont {H.~M.}\ \bibnamefont
  {Aktulga}}, \bibinfo {author} {\bibfnamefont {R.}~\bibnamefont {Berger}},
  \bibinfo {author} {\bibfnamefont {D.~S.}\ \bibnamefont {Bolintineanu}},
  \bibinfo {author} {\bibfnamefont {W.~M.}\ \bibnamefont {Brown}}, \bibinfo
  {author} {\bibfnamefont {P.~S.}\ \bibnamefont {Crozier}}, \bibinfo {author}
  {\bibfnamefont {P.~J.}\ \bibnamefont {in~'t Veld}}, \bibinfo {author}
  {\bibfnamefont {A.}~\bibnamefont {Kohlmeyer}}, \bibinfo {author}
  {\bibfnamefont {S.~G.}\ \bibnamefont {Moore}}, \bibinfo {author}
  {\bibfnamefont {T.~D.}\ \bibnamefont {Nguyen}}, \bibinfo {author}
  {\bibfnamefont {R.}~\bibnamefont {Shan}}, \bibinfo {author} {\bibfnamefont
  {M.~J.}\ \bibnamefont {Stevens}}, \bibinfo {author} {\bibfnamefont
  {J.}~\bibnamefont {Tranchida}}, \bibinfo {author} {\bibfnamefont
  {C.}~\bibnamefont {Trott}},\ and\ \bibinfo {author} {\bibfnamefont {S.~J.}\
  \bibnamefont {Plimpton}},\ }\bibfield  {title} {\bibinfo {title} {{LAMMPS} -
  a flexible simulation tool for particle-based materials modeling at the
  atomic, meso, and continuum scales},\ }\href@noop {} {\bibfield  {journal}
  {\bibinfo  {journal} {Comp. Phys. Comm.}\ }\textbf {\bibinfo {volume}
  {271}},\ \bibinfo {pages} {108171} (\bibinfo {year} {2022})}\BibitemShut
  {NoStop}%
\bibitem [{\citenamefont {Hirel}(2015)}]{hirel2015atomsk}%
  \BibitemOpen
  \bibfield  {author} {\bibinfo {author} {\bibfnamefont {P.}~\bibnamefont
  {Hirel}},\ }\bibfield  {title} {\bibinfo {title} {Atomsk: A tool for
  manipulating and converting atomic data files},\ }\href@noop {} {\bibfield
  {journal} {\bibinfo  {journal} {Computer Physics Communications}\ }\textbf
  {\bibinfo {volume} {197}},\ \bibinfo {pages} {212} (\bibinfo {year}
  {2015})}\BibitemShut {NoStop}%
\bibitem [{\citenamefont {Hoover}(1985)}]{hoover1985canonical}%
  \BibitemOpen
  \bibfield  {author} {\bibinfo {author} {\bibfnamefont {W.~G.}\ \bibnamefont
  {Hoover}},\ }\bibfield  {title} {\bibinfo {title} {Canonical dynamics:
  Equilibrium phase-space distributions},\ }\href@noop {} {\bibfield  {journal}
  {\bibinfo  {journal} {Physical review A}\ }\textbf {\bibinfo {volume} {31}},\
  \bibinfo {pages} {1695} (\bibinfo {year} {1985})}\BibitemShut {NoStop}%
\bibitem [{\citenamefont {Sadigh}\ \emph {et~al.}(2012)\citenamefont {Sadigh},
  \citenamefont {Erhart}, \citenamefont {Stukowski}, \citenamefont {Caro},
  \citenamefont {Martinez},\ and\ \citenamefont
  {Zepeda-Ruiz}}]{sadigh2012scalable}%
  \BibitemOpen
  \bibfield  {author} {\bibinfo {author} {\bibfnamefont {B.}~\bibnamefont
  {Sadigh}}, \bibinfo {author} {\bibfnamefont {P.}~\bibnamefont {Erhart}},
  \bibinfo {author} {\bibfnamefont {A.}~\bibnamefont {Stukowski}}, \bibinfo
  {author} {\bibfnamefont {A.}~\bibnamefont {Caro}}, \bibinfo {author}
  {\bibfnamefont {E.}~\bibnamefont {Martinez}},\ and\ \bibinfo {author}
  {\bibfnamefont {L.}~\bibnamefont {Zepeda-Ruiz}},\ }\bibfield  {title}
  {\bibinfo {title} {Scalable parallel {M}onte {C}arlo algorithm for atomistic
  simulations of precipitation in alloys},\ }\href@noop {} {\bibfield
  {journal} {\bibinfo  {journal} {Physical Review B}\ }\textbf {\bibinfo
  {volume} {85}},\ \bibinfo {pages} {184203} (\bibinfo {year}
  {2012})}\BibitemShut {NoStop}%
\bibitem [{\citenamefont {Stukowski}(2009)}]{stukowski2009visualization}%
  \BibitemOpen
  \bibfield  {author} {\bibinfo {author} {\bibfnamefont {A.}~\bibnamefont
  {Stukowski}},\ }\bibfield  {title} {\bibinfo {title} {Visualization and
  analysis of atomistic simulation data with {OVITO}--the {O}pen
  {V}isualization {T}ool},\ }\href@noop {} {\bibfield  {journal} {\bibinfo
  {journal} {Modelling and Simulation in Materials Science and Engineering}\
  }\textbf {\bibinfo {volume} {18}},\ \bibinfo {pages} {015012} (\bibinfo
  {year} {2009})}\BibitemShut {NoStop}%
\bibitem [{\citenamefont {Nowotny}(1946)}]{nowotny1946kristallstrukturen}%
  \BibitemOpen
  \bibfield  {author} {\bibinfo {author} {\bibfnamefont {H.}~\bibnamefont
  {Nowotny}},\ }\bibfield  {title} {\bibinfo {title} {Die {K}ristallstrukturen
  von {Z}n9{T}h, {C}d2{C}a und ({A}g, {M}g)2{C}a},\ }\href@noop {} {\bibfield
  {journal} {\bibinfo  {journal} {International Journal of Materials Research}\
  }\textbf {\bibinfo {volume} {37}},\ \bibinfo {pages} {31} (\bibinfo {year}
  {1946})}\BibitemShut {NoStop}%
\bibitem [{\citenamefont {Arr{\'o}yave}\ and\ \citenamefont
  {Liu}(2006)}]{arroyave2006intermetallics}%
  \BibitemOpen
  \bibfield  {author} {\bibinfo {author} {\bibfnamefont {R.}~\bibnamefont
  {Arr{\'o}yave}}\ and\ \bibinfo {author} {\bibfnamefont {Z.-K.}\ \bibnamefont
  {Liu}},\ }\bibfield  {title} {\bibinfo {title} {Intermetallics in the
  {M}g-{C}a-{S}n ternary system: {S}tructural, vibrational, and thermodynamic
  properties from first principles},\ }\href@noop {} {\bibfield  {journal}
  {\bibinfo  {journal} {Physical Review B}\ }\textbf {\bibinfo {volume} {74}},\
  \bibinfo {pages} {174118} (\bibinfo {year} {2006})}\BibitemShut {NoStop}%
\bibitem [{\citenamefont {Kammerer}\ \emph {et~al.}(2015)\citenamefont
  {Kammerer}, \citenamefont {Behdad}, \citenamefont {Zhou}, \citenamefont
  {Betancor}, \citenamefont {Gonzalez}, \citenamefont {Boesl},\ and\
  \citenamefont {Sohn}}]{kammerer2015diffusion}%
  \BibitemOpen
  \bibfield  {author} {\bibinfo {author} {\bibfnamefont {C.}~\bibnamefont
  {Kammerer}}, \bibinfo {author} {\bibfnamefont {S.}~\bibnamefont {Behdad}},
  \bibinfo {author} {\bibfnamefont {L.}~\bibnamefont {Zhou}}, \bibinfo {author}
  {\bibfnamefont {F.}~\bibnamefont {Betancor}}, \bibinfo {author}
  {\bibfnamefont {M.}~\bibnamefont {Gonzalez}}, \bibinfo {author}
  {\bibfnamefont {B.}~\bibnamefont {Boesl}},\ and\ \bibinfo {author}
  {\bibfnamefont {Y.}~\bibnamefont {Sohn}},\ }\bibfield  {title} {\bibinfo
  {title} {Diffusion kinetics, mechanical properties, and crystallographic
  characterization of intermetallic compounds in the {M}g-{Z}n binary system},\
  }\href@noop {} {\bibfield  {journal} {\bibinfo  {journal} {Intermetallics}\
  }\textbf {\bibinfo {volume} {67}},\ \bibinfo {pages} {145} (\bibinfo {year}
  {2015})}\BibitemShut {NoStop}%
\bibitem [{\citenamefont {Jain}\ \emph {et~al.}(2013)\citenamefont {Jain},
  \citenamefont {Ong}, \citenamefont {Hautier}, \citenamefont {Chen},
  \citenamefont {Richards}, \citenamefont {Dacek}, \citenamefont {Cholia},
  \citenamefont {Gunter}, \citenamefont {Skinner}, \citenamefont {Ceder} \emph
  {et~al.}}]{jain2013commentary}%
  \BibitemOpen
  \bibfield  {author} {\bibinfo {author} {\bibfnamefont {A.}~\bibnamefont
  {Jain}}, \bibinfo {author} {\bibfnamefont {S.~P.}\ \bibnamefont {Ong}},
  \bibinfo {author} {\bibfnamefont {G.}~\bibnamefont {Hautier}}, \bibinfo
  {author} {\bibfnamefont {W.}~\bibnamefont {Chen}}, \bibinfo {author}
  {\bibfnamefont {W.~D.}\ \bibnamefont {Richards}}, \bibinfo {author}
  {\bibfnamefont {S.}~\bibnamefont {Dacek}}, \bibinfo {author} {\bibfnamefont
  {S.}~\bibnamefont {Cholia}}, \bibinfo {author} {\bibfnamefont
  {D.}~\bibnamefont {Gunter}}, \bibinfo {author} {\bibfnamefont
  {D.}~\bibnamefont {Skinner}}, \bibinfo {author} {\bibfnamefont
  {G.}~\bibnamefont {Ceder}}, \emph {et~al.},\ }\bibfield  {title} {\bibinfo
  {title} {Commentary: The materials project: A materials genome approach to
  accelerating materials innovation},\ }\href@noop {} {\bibfield  {journal}
  {\bibinfo  {journal} {APL materials}\ }\textbf {\bibinfo {volume} {1}},\
  \bibinfo {pages} {011002} (\bibinfo {year} {2013})}\BibitemShut {NoStop}%
\bibitem [{\citenamefont {Sumer}\ and\ \citenamefont
  {Smith}(1962)}]{sumer1962elastic}%
  \BibitemOpen
  \bibfield  {author} {\bibinfo {author} {\bibfnamefont {A.}~\bibnamefont
  {Sumer}}\ and\ \bibinfo {author} {\bibfnamefont {J.}~\bibnamefont {Smith}},\
  }\bibfield  {title} {\bibinfo {title} {Elastic constants of single-crystal
  {C}a{M}g2},\ }\href@noop {} {\bibfield  {journal} {\bibinfo  {journal}
  {Journal of Applied Physics}\ }\textbf {\bibinfo {volume} {33}},\ \bibinfo
  {pages} {2283} (\bibinfo {year} {1962})}\BibitemShut {NoStop}%
\bibitem [{\citenamefont {Ganeshan}\ \emph {et~al.}(2009)\citenamefont
  {Ganeshan}, \citenamefont {Shang}, \citenamefont {Zhang}, \citenamefont
  {Wang}, \citenamefont {Mantina},\ and\ \citenamefont
  {Liu}}]{ganeshan2009elastic}%
  \BibitemOpen
  \bibfield  {author} {\bibinfo {author} {\bibfnamefont {S.}~\bibnamefont
  {Ganeshan}}, \bibinfo {author} {\bibfnamefont {S.}~\bibnamefont {Shang}},
  \bibinfo {author} {\bibfnamefont {H.}~\bibnamefont {Zhang}}, \bibinfo
  {author} {\bibfnamefont {Y.}~\bibnamefont {Wang}}, \bibinfo {author}
  {\bibfnamefont {M.}~\bibnamefont {Mantina}},\ and\ \bibinfo {author}
  {\bibfnamefont {Z.}~\bibnamefont {Liu}},\ }\bibfield  {title} {\bibinfo
  {title} {Elastic constants of binary {M}g compounds from first-principles
  calculations},\ }\href@noop {} {\bibfield  {journal} {\bibinfo  {journal}
  {Intermetallics}\ }\textbf {\bibinfo {volume} {17}},\ \bibinfo {pages} {313}
  (\bibinfo {year} {2009})}\BibitemShut {NoStop}%
\bibitem [{\citenamefont {Seidenkranz}\ and\ \citenamefont
  {Hegenbarth}(1976)}]{seidenkranz1976single}%
  \BibitemOpen
  \bibfield  {author} {\bibinfo {author} {\bibfnamefont {T.}~\bibnamefont
  {Seidenkranz}}\ and\ \bibinfo {author} {\bibfnamefont {E.}~\bibnamefont
  {Hegenbarth}},\ }\bibfield  {title} {\bibinfo {title} {Single-crystal elastic
  constants of {M}g{Z}n2 in the temperature range from 4.2 to 300 {K}},\
  }\href@noop {} {\bibfield  {journal} {\bibinfo  {journal} {Physica Status
  Solidi (a)}\ }\textbf {\bibinfo {volume} {33}},\ \bibinfo {pages} {205}
  (\bibinfo {year} {1976})}\BibitemShut {NoStop}%
\bibitem [{\citenamefont {Tehranchi}\ \emph {et~al.}(2022)\citenamefont
  {Tehranchi}, \citenamefont {Hickel},\ and\ \citenamefont
  {Neugebauer}}]{Tehranchi}%
  \BibitemOpen
  \bibfield  {author} {\bibinfo {author} {\bibfnamefont {A.}~\bibnamefont
  {Tehranchi}}, \bibinfo {author} {\bibfnamefont {T.}~\bibnamefont {Hickel}},\
  and\ \bibinfo {author} {\bibfnamefont {J.}~\bibnamefont {Neugebauer}},\
  }\bibfield  {title} {\bibinfo {title} {private communication},\ }\href@noop
  {} {\  (\bibinfo {year} {2022})}\BibitemShut {NoStop}%
\bibitem [{\citenamefont {Rennert}\ and\ \citenamefont
  {Radwan}(1977)}]{rennert1977structural}%
  \BibitemOpen
  \bibfield  {author} {\bibinfo {author} {\bibfnamefont {P.}~\bibnamefont
  {Rennert}}\ and\ \bibinfo {author} {\bibfnamefont {A.}~\bibnamefont
  {Radwan}},\ }\bibfield  {title} {\bibinfo {title} {Structural investigation
  of the laves phase {M}g{Z}n2 with model potential calculations},\ }\href@noop
  {} {\bibfield  {journal} {\bibinfo  {journal} {Physica Status Solidi (b)}\
  }\textbf {\bibinfo {volume} {79}},\ \bibinfo {pages} {167} (\bibinfo {year}
  {1977})}\BibitemShut {NoStop}%
\bibitem [{\citenamefont {Ma}\ \emph {et~al.}(2013)\citenamefont {Ma},
  \citenamefont {Fan}, \citenamefont {Tang}, \citenamefont {Peng},\ and\
  \citenamefont {Ding}}]{ma2013ab}%
  \BibitemOpen
  \bibfield  {author} {\bibinfo {author} {\bibfnamefont {L.}~\bibnamefont
  {Ma}}, \bibinfo {author} {\bibfnamefont {T.-W.}\ \bibnamefont {Fan}},
  \bibinfo {author} {\bibfnamefont {B.-Y.}\ \bibnamefont {Tang}}, \bibinfo
  {author} {\bibfnamefont {L.-M.}\ \bibnamefont {Peng}},\ and\ \bibinfo
  {author} {\bibfnamefont {W.-j.}\ \bibnamefont {Ding}},\ }\bibfield  {title}
  {\bibinfo {title} {Ab initio study of {I}2 and {T}2 stacking faults in {C}14
  {L}aves phase {M}g{Z}n2},\ }\href@noop {} {\bibfield  {journal} {\bibinfo
  {journal} {The European Physical Journal B}\ }\textbf {\bibinfo {volume}
  {86}},\ \bibinfo {pages} {1} (\bibinfo {year} {2013})}\BibitemShut {NoStop}%
\end{thebibliography}%

\end{document}